%% file: main.tex
\documentclass[10pt]{article}

\usepackage[T1]{fontenc}
\usepackage[utf8]{inputenc}

\usepackage{newtxtext}
\usepackage{amsmath,amssymb,amsfonts}
\usepackage{newtxmath}

\usepackage[margin=1in]{geometry}
\usepackage{microtype}
\usepackage{titlesec}
\usepackage{setspace}
\usepackage{float}

\titleformat{\section}{\large\bfseries}{\thesection.}{0.6em}{}
\titleformat{\subsection}{\normalsize\bfseries}{\thesubsection}{0.6em}{}
\titleformat{\subsubsection}{\normalsize\itshape}{\thesubsubsection}{0.6em}{}
\titlespacing*{\section}{0pt}{2.2ex plus .4ex minus .2ex}{1.0ex}
\titlespacing*{\subsection}{0pt}{1.7ex plus .3ex minus .2ex}{0.7ex}
\titlespacing*{\subsubsection}{0pt}{1.2ex plus .2ex minus .1ex}{0.5ex}

\usepackage{mathtools}

\usepackage{graphicx}
\usepackage{caption}
\usepackage{subcaption}
\usepackage{array}
\usepackage{booktabs}
\usepackage{multirow}
\usepackage[table]{xcolor}

\usepackage{tikz}
\usetikzlibrary{positioning,calc,arrows.meta,shapes.geometric,shapes.misc,fit,backgrounds}

\usepackage[ruled,vlined,linesnumbered]{algorithm2e}
\usepackage{enumitem}

\usepackage{cite}
\usepackage{xurl}
\usepackage{comment}
\usepackage{siunitx}
\usepackage[hidelinks]{hyperref}
\hypersetup{
  pdftitle={Prefix Visibility Ratio: Quantifying Partial Observability in Carpet Bombing Detection},
  pdfauthor={Farid Nait-Abdesselam, Laaziz Lahlou, Doha Bentaoussy, Hugo Lege, Eve Lin, Nadjia Kara}
}

\newcolumntype{L}[1]{>{\raggedright\arraybackslash}p{#1}}
\newcolumntype{C}[1]{>{\centering\arraybackslash}p{#1}}
\def\BibTeX{{\rm B\kern-.05em{\sc i\kern-.025em b}\kern-.08em T\kern-.1667em\lower.7ex\hbox{E}\kern-.125emX}}

\begin{document}

\title{\textbf{Prefix Visibility Ratio: Quantifying Partial Observability in Carpet Bombing Detection}}

\author{%
\begin{tabular}{ccc}
\small Farid Naït-Abdesselam$^{1}$ &
\small Laaziz Lahlou$^{2}$ &
\small Doha Bentaoussy$^{1}$ \\
\small Hugo Lege$^{1}$ &
\small Eve Lin$^{1}$ &
\small Nadjia Kara$^{2}$
\end{tabular}
\\[1.8em]
\small $^{1}$Université Paris Cité, LIPADE, F-75006 Paris, France\\[-.3em]
\small $^{2}$École de Technologie Supérieure (ÉTS), Montréal, Canada
}
\date{\small October 2026}

\maketitle

\begin{abstract}
Carpet bombing attacks distribute malicious traffic across many destinations within a network prefix, while operational monitoring commonly provides destination-specific telemetry for only a subset of those targets. This creates an attack-conditional observability problem that is not captured by address utilization or nominal monitoring coverage. We introduce the \emph{Prefix Visibility Ratio} (PVR), defined as the fraction of attacked destinations for which the telemetry required by the host-level evidence plane is available, and argue that PVR should be treated as an explicit evaluation dimension for carpet bombing detection. We then present VANTA (Visibility-Aware Network Threat Analysis) as one concrete way to operationalize this principle. VANTA combines host-level and prefix-level anomaly scores, estimates PVR online from router-side destination evidence and the monitored-destination inventory, and adjusts evidence fusion according to both estimated visibility and estimator confidence. Across a validated campaign of 2,160 Mininet runs, VANTA achieves 99.18\% precision, 93.46\% recall, a global F1-score of 96.24\%, and a 1.54\% false-positive rate. In a separate matched-observability study using task-adapted feature-level reference implementations, VANTA reaches 95.89\% pooled F1 and 92.39\% recall for configured $PVR\leq10\%$, with its largest advantage appearing at configured $PVR=1\%$. These reference comparisons are intended to isolate observability effects under a common binary detection task rather than to reproduce the original end-to-end objectives of the referenced systems. The results establish attacked-destination visibility as a measurable factor that can materially change detector behavior and should be reported explicitly in future carpet bombing evaluations.
\end{abstract}

\noindent\textbf{Keywords:} carpet bombing, distributed denial-of-service, network observability, Prefix Visibility Ratio, partial telemetry, visibility-aware detection.

\section{Introduction}
\label{sec:intro}

Carpet bombing is a distributed denial-of-service (DDoS) strategy in which malicious traffic is spread across many addresses belonging to the same network prefix. The objective is not necessarily to overload any single host. Instead, the aggregate load threatens shared access links, aggregation devices, and upstream connectivity while the traffic received by each destination can remain below conventional host-level thresholds~\cite{mirkovic2004taxonomy,zargar2013survey,heinrich2021carpetbombing,huici2007edge,wagner2021united}. This spatial dispersion makes cross-destination reasoning important, but it also exposes a less studied problem: \textit{the defender may not obtain destination-specific telemetry from most of the addresses selected by the attacker}.

Partial observability can arise for several reasons. Operational prefixes contain inactive or unassigned addresses, transient clients, hosts without monitoring agents, and services that expose different levels of telemetry~\cite{richter2015surveying,caida,liu2016middlepolice,hu2025sistar}. Network-side collectors can still observe aggregate or per-destination traffic at gateways, but this does not imply that rich host-level evidence is available for every attacked destination. Consequently, a detector may receive strong prefix-wide evidence while observing only a small and potentially unrepresentative subset of the attacked hosts.

This distinction is important because \emph{monitoring coverage} and \emph{attack visibility} are not the same quantity. A network can instrument many active hosts while an attacker happens to target mainly unmonitored addresses. Conversely, modest monitoring coverage can provide substantial attack evidence when the attacked set overlaps strongly with the monitored set. The relevant quantity is therefore conditional on both the defender's deployment and the attacker's selected destinations.

We formalize this attack-conditional overlap through the \emph{Prefix Visibility Ratio} (PVR). For an attacked set $A_t$ and a destination-observable set $O_t$ during observation window $t$, PVR is $|A_t\cap O_t|/|A_t|$. Here, $O_t$ specifically denotes destinations providing the telemetry required by the host-level evidence plane. Gateway measurements, including per-destination ingress rates, remain complementary network-level evidence and do not by themselves place a destination in $O_t$. PVR therefore measures visibility relative to this declared evidence plane. This makes it distinct from the Address Utilization Ratio (AUR), which describes how much of the prefix is active, and from nominal monitoring coverage, which describes how much of the active address space is instrumented independently of the attack.

To study how an explicit visibility variable can be used by a detector, we present \textbf{VANTA} (\emph{Visibility-Aware Network Threat Analysis}). VANTA combines host-level anomaly evidence with prefix-level measurements, estimates PVR online without consulting attack ground truth, and adapts the host/prefix fusion weight according to the estimated visibility and its confidence. The estimator uses router-side destination-rate anomalies to infer the attacked set when such information is available, intersects that estimate with the monitored-destination inventory, and applies temporal smoothing. When visibility is low or uncertain, VANTA shifts weight toward prefix-level evidence instead of treating missing host observations as benign.

Our evaluation has two purposes. First, the primary 2,160-run Mininet campaign characterizes VANTA across a controlled AUR--PVR--attack-extent grid, including detection performance, delay, PVR-estimation error, and complete-miss cases. Second, a separate comparative campaign studies three transparent and feature-level reference adaptations derived from prior work~\cite{netradar,flowlens,whisper} under matched destination observability. These adaptations are useful for studying sensitivity to missing destination evidence, but they are \emph{not} presented as end-to-end reproductions of the original systems or as a replacement for their original tasks. In particular, NetRadar's original objective includes victim localization, whereas our comparative task is binary prefix-level attack detection.

The main contributions are:
\begin{enumerate}[leftmargin=*,itemsep=2pt,topsep=4pt]
\item We introduce PVR as an explicit and attack-conditional metric for attacked-destination observability and distinguish it formally from AUR and nominal monitoring coverage.
\item We present VANTA, a visibility-aware detection framework with an online PVR estimator, independent host- and prefix-level anomaly detectors, and confidence-aware adaptive fusion.
\item We document the estimator and decision pipeline in implementation-level detail, including the exact rounding convention for effective PVR and the thresholds used by the validated campaign.
\item We evaluate VANTA over 2,160 Mininet runs spanning 72 AUR--PVR--attack-extent configurations, and we separately use task-adapted reference implementations to study how matched observability changes binary detection behavior.
To support reproducibility, we publicly release the implementation, experimental configurations, analysis scripts, and validated data used in this study.\footnote{\url{https://doi.org/10.5281/zenodo.23534795}}
\end{enumerate}

The paper first reviews related work (Section~\ref{sec:related_work}), formalizes the observability problem and PVR (Sections~\ref{sec:visibility_problem} and~\ref{sec:pvr}), states the threat model (Section~\ref{sec:threat_model}), and describes VANTA (Section~\ref{sec:vanta}). Sections~\ref{sec:methodology} and~\ref{sec:vanta_evaluation} present the evaluation and results, followed by interpretation, limitations, and future directions in Section~\ref{sec:discussion}.

\section{Related Work}
\label{sec:related_work}

\subsection{Carpet Bombing Detection}

Carpet bombing attacks distribute malicious traffic across multiple destination addresses belonging to a common network prefix. Instead of concentrating traffic on a single victim, the attacker spreads the load over a larger destination set, thereby reducing the traffic intensity and anomaly evidence observable at any individual host while still threatening shared access links, routers, or upstream infrastructure~\cite{heinrich2021carpetbombing,mirkovic2004taxonomy,zargar2013survey}. This behavior makes carpet bombing fundamentally different from conventional single-victim volumetric DDoS attacks, for which host-level rate thresholds or per-destination anomaly indicators are often sufficient.

Measurement studies have documented a broad range of distributed denial-of-service behaviors, including reflection-based, multiprotocol, low-rate, and spatially distributed attacks~\cite{heinrich2021carpetbombing,bailey2005network,moore2006inferring,sommese2022investigating,nawrocki2021far}. These studies show that attackers can vary both traffic composition and destination distribution, complicating defenses that rely on strong anomalies at individual endpoints. Earlier network-wide defenses therefore exploit aggregate traffic statistics, sketches, entropy changes, congestion signals, or anomaly-detection mechanisms to identify deviations that may remain weak when considered at the level of a single destination~\cite{lakhina2004diagnosing,barford2002signal,krishnamurthy2003sketch,alcoz2022aggregate}. Such techniques are naturally less dependent on complete end-host instrumentation because they operate primarily on traffic observed at network aggregation points.

However, aggregate visibility and attacked-destination visibility are not equivalent. A border router may observe substantial traffic entering a prefix while only a small subset of the destinations selected by the attacker exposes destination-specific telemetry to the detection system. Conversely, extensive monitoring deployment does not guarantee that monitored hosts overlap substantially with the attacker-selected destination set. Most prior work characterizes the traffic available at the measurement point, but does not explicitly parameterize this overlap as an independent dimension of the detection problem.

NetRadar is the closest dedicated carpet bombing system to our setting~\cite{netradar}. It combines gateway-side traffic information with server-side feedback and permutation-equivariant cross-server analysis in order to exploit weak evidence distributed across multiple destinations. Its architecture demonstrates the value of correlating evidence across servers rather than treating each destination independently. However, its end-to-end task includes identifying or localizing attacked servers at a given time, whereas VANTA focuses on determining whether the monitored prefix is experiencing a carpet bombing attack under incomplete destination observability.

This distinction is important when interpreting comparative results. Failure to identify a particular attacked server does not necessarily imply failure to recognize that an attack is occurring somewhere within the prefix. Conversely, reliable prefix-level detection does not imply accurate victim localization. For this reason, we do not present the VANTA evaluation as an exact end-to-end reproduction of NetRadar. Instead, NetRadar-S is used as a transparent and feature-level adaptation that preserves representative traffic, server-side, and cross-destination evidence while operating on the same binary prefix-level detection task considered by VANTA. The objective of this comparison is therefore to study how detection evidence changes as attacked-destination observability decreases, rather than to claim equivalence with NetRadar's original localization objective.

\subsection{Flow and Frequency-Domain Traffic Classification}

A second line of work detects malicious traffic by learning discriminative representations from packet- or flow-level observations. FlowLens derives compact statistical representations from packet-size and inter-arrival distributions and applies supervised learning to support efficient traffic classification~\cite{flowlens}. Its design emphasizes feature reduction and efficient classification while retaining traffic characteristics that distinguish malicious from benign flows.

Whisper approaches malicious-traffic detection from a different representation perspective~\cite{whisper}. It encodes packet size, protocol information, and inter-arrival timing into temporal sequences and transforms them into frequency-domain features. This representation is designed to capture periodicity and structural patterns that may be less apparent in raw packet sequences. NetRadar subsequently adapts both FlowLens and Whisper within its carpet bombing evaluation framework~\cite{netradar}, demonstrating that flow-level and spectral features can provide useful evidence in distributed attack scenarios.

These approaches remain highly relevant to our study because they represent strong traffic-based detection strategies that do not inherently require the same form of host instrumentation as VANTA. However, their original objectives concern malicious-flow or traffic-pattern classification rather than the explicit modeling of attacked-destination observability. Classification performance is normally evaluated on the traffic records made available to the detector. If traffic associated with a destination is absent from the detector's observable input, even a highly accurate per-flow classifier cannot directly recover destination-specific evidence that was never collected.

This distinction becomes particularly important for carpet bombing attacks, where the attack signal is intentionally dispersed across a destination set. A detector may perform extremely well when packet or flow features for all attacked destinations are available, yet experience a substantial reduction in usable evidence when only a small subset of those destinations is observable. The resulting degradation is not necessarily a weakness of the underlying classifier, it is a consequence of the measurement conditions under which the classifier operates.

Accordingly, our FlowLens-S and Whisper-S implementations retain representative feature families and supervised-classification structures while placing all methods within a common binary prefix-level detection setting. As with NetRadar-S, they should be interpreted as \emph{method-faithful reference adaptations}, rather than exact reproductions of the original end-to-end systems. This distinction allows us to separate the question of traffic-classification capability from the question addressed in this work: \textit{how the availability of attacked-destination evidence affects detection under partial observability}.

\subsection{Partial Observability and Missing Telemetry}

Incomplete monitoring has long been recognized as a fundamental challenge in network measurement, tomography, anomaly detection, and traffic inference~\cite{lakhina2004diagnosing,garcia2009anomaly,bhuyan2014network,sommer2010outside,estan2002newdirections,claise2008ipfix}. Operational monitoring infrastructures rarely provide a complete view of all devices, flows, or network events. Measurements may be limited by sampling, deployment cost, instrumentation coverage, routing asymmetry, inactive hosts, telemetry loss, or the administrative boundaries of the monitored infrastructure.

A substantial body of work therefore investigates how to operate under incomplete measurements. Sketch-based mechanisms compress large traffic streams while preserving information useful for anomaly detection~\cite{krishnamurthy2003sketch}. Sampling and traffic-inference techniques reduce measurement overhead while attempting to retain representative network information. Other approaches use coordinated router instrumentation or programmable network infrastructure to improve the resolution and usefulness of available measurements~\cite{kompella2014every,feamster2014road}. These techniques address important forms of partial information, but they primarily concern how much network state can be measured or reconstructed.

The observability problem considered in this paper is related but different. Nominal monitoring coverage describes what fraction of a network is instrumented, while the relevant quantity during an attack depends on the intersection between that instrumentation and the destination set actually selected by the attacker. Let $O$ denote the destinations capable of providing destination-specific evidence and $A$ the destinations targeted during an attack. A network may have extensive monitoring coverage while still exposing little useful host-level evidence if $A$ lies predominantly outside $O$. Conversely, relatively sparse monitoring can provide substantial attack evidence when the attacker happens to target destinations contained in the monitored subset.

This attack-conditioned view motivates the Prefix Visibility Ratio (PVR), which measures the fraction of attacked destinations observable to the defender. PVR therefore does not attempt to replace existing notions of telemetry availability, sampling coverage, or address-space utilization. Instead, it introduces a complementary quantity that captures the interaction between attacker behavior and defender observability.

This distinction also separates PVR from the Address Utilization Ratio (AUR). AUR describes how much of the target address space is active, independently of the destinations selected by the attacker. PVR instead depends on the attacked set itself and may vary between attacks against the same network without any change in the underlying monitoring deployment. Thus, two networks with identical AUR and nominal monitoring coverage can expose substantially different amounts of useful evidence depending on which destinations are attacked.

Our contribution is consequently not a new general-purpose missing-data reconstruction technique. Rather, we make attacked-destination observability an explicit property of the threat model, define a metric for quantifying it, estimate it online, and use it both as an evaluation axis and as an input to adaptive detection. This perspective complements prior work on network measurement and missing telemetry by asking a more attack-specific question: \emph{given the destinations selected by the adversary, how much of the relevant attack surface is actually observable to the detector?}

\section{The Prefix Visibility Problem}
\label{sec:visibility_problem}

Let $P$ denote the set of usable addresses in a targeted network prefix, $U_t\subseteq P$ the addresses that are active during observation window $t$, $O_t\subseteq U_t$ the active destinations that expose destination-specific telemetry to the defender, and $A_t\subseteq P$ the destinations receiving attack traffic. The sets $U_t$, $O_t$, and $A_t$ need not coincide and may evolve over time as hosts become active or inactive, monitoring agents appear or disappear, and the attacker changes its destination set.

This distinction is particularly important for carpet bombing attacks. The attack objective is typically to create aggregate pressure on shared network infrastructure rather than to overload one specific endpoint. Consequently, a destination can contribute to the attack's network-level impact even when it provides no host-level evidence to the defender. An inactive or uninstrumented address may therefore receive malicious traffic that contributes to congestion while remaining absent from destination-specific telemetry.

Prefix-level measurements collected at a gateway or aggregation point are deliberately kept separate from $O_t$. Such measurements can include aggregate traffic volume, destination dispersion, source diversity, flow fan-out, or traffic entropy, and remain available even when a particular destination exposes no host telemetry. In contrast, membership in $O_t$ indicates that the defender can obtain destination-specific evidence, such as local traffic counters, connection statistics, protocol distributions, or host resource measurements. This separation allows us to distinguish two fundamentally different sources of evidence: network-wide observations that characterize the prefix as a whole and host-specific observations that characterize individual destinations.

\subsection{Address Utilization and Nominal Monitoring Coverage}

Operational prefixes are rarely populated uniformly. Enterprise, campus, cloud, and service-provider address spaces may contain permanently active systems, transient clients, reserved addresses, temporarily unused allocations, and inactive addresses. We characterize the fraction of the usable prefix that is active during observation window $t$ through the Address Utilization Ratio
\begin{equation}
AUR_t=\frac{|U_t|}{|P|}.
\label{eq:aur}
\end{equation}

$AUR_t$ captures how densely the target prefix is populated, but it contains no information about the defender's instrumentation or about the attack itself. Two prefixes can therefore have the same $AUR_t$ while exposing very different amounts of telemetry. Similarly, changes in $AUR_t$ may occur independently of the attack, for example because of ordinary client churn or dynamic address allocation.

To separate utilization from monitoring deployment, we define nominal destination monitoring coverage as
\begin{equation}
MC_t=\frac{|O_t|}{|U_t|},
\label{eq:monitoring_coverage}
\end{equation}
for $|U_t|>0$. $MC_t$ describes the fraction of active destinations from which destination-specific evidence can potentially be collected. It is primarily a property of the monitoring infrastructure and may depend on factors such as agent deployment, host ownership, telemetry availability, administrative policy, or collection cost.

Importantly, neither $AUR_t$ nor $MC_t$ indicates how much of a particular attack is visible. A detector may operate in a highly utilized prefix with broad monitoring coverage and still receive little destination-specific evidence if the attacker predominantly targets addresses outside $O_t$. Conversely, a deployment with modest nominal monitoring coverage may expose strong attack evidence if the attacked destinations happen to lie largely inside the monitored subset.

This distinction shows why monitoring coverage alone is insufficient for characterizing the information available to a carpet bombing detector. $MC_t$ answers the question ``How much of the active prefix can be monitored?'', whereas attack detection depends on the different question ``How much of the attacker-selected destination set is actually observable?''

\subsection{Attack-Conditional Visibility}

Carpet bombing creates a structural mismatch between where damage accumulates and where detailed evidence is collected. The attacker distributes traffic across many destinations, while the defender may observe host-specific behavior for only a subset of them. Nevertheless, traffic delivered to unmonitored, inactive, or otherwise unobservable destinations can still consume bandwidth and forwarding capacity at shared links and aggregation devices. As a result, substantial network-level impact may coexist with weak host-level evidence.

The amount of useful host telemetry therefore depends not only on how many destinations are monitored, but on the overlap between the monitored set $O_t$ and the attack target set $A_t$. This overlap is attack-dependent. Even if $O_t$ remains fixed, two attacks against the same prefix can provide very different levels of destination-specific evidence simply because they target different address subsets.

For example, consider two deployments with identical $AUR_t$ and $MC_t$. In the first, most attacked destinations belong to $O_t$, so host-level observations provide multiple corroborating signals. In the second, the attacker targets primarily destinations outside $O_t$. The aggregate load delivered to the prefix may be comparable in both cases, yet the second attack produces substantially less host-level evidence. From the perspective of a detector that relies on cross-host observations, these two situations are fundamentally different despite having identical utilization and nominal monitoring coverage.

The same phenomenon can arise dynamically. An attacker may rotate its target addresses across observation windows while maintaining a similar aggregate rate. Such movement can change the overlap between $A_t$ and $O_t$ without any modification to the monitoring infrastructure. Consequently, the effective observability of the attack may vary over time even when $AUR_t$ and $MC_t$ remain nearly constant.

\begin{figure}[t]
\centering
\resizebox{0.5\textwidth}{!}{%
\begin{tikzpicture}[
    x=3.6mm,y=3.6mm,
    cell/.style={rectangle,draw=black!70,line width=0.25pt,minimum width=2.8mm,minimum height=2.8mm,inner sep=0pt,outer sep=0pt},
    empty/.style={cell,fill=white},
    attacked/.style={cell,fill=red!78},
    observed/.style={cell,fill=green!55!black!45},
    arrow/.style={->,thick,>=Stealth}
]
\foreach \x in {0,...,15}{\foreach \y in {0,...,15}{\node[attacked] at (\x,-\y) {};}}
\node[empty] at (0,0) {}; \node[empty] at (15,-15) {};
\begin{scope}[xshift=7.2cm]
\foreach \x in {0,...,15}{\foreach \y in {0,...,15}{\node[empty] at (\x,-\y) {};}}
\foreach \x/\y in {3/0,9/0,4/1,14/1,0/2,12/2,5/3,2/4,6/4,8/4,6/5,3/6,12/7,2/8,6/8,13/8,8/9,9/10,1/11,15/11,1/12,4/12,7/12,15/12,2/13,13/13,2/14,12/14,14/14}{\node[observed] at (\x,-\y) {};}
\end{scope}
\draw[arrow] (16.5,-7.5) -- (18.4,-7.5);
\node[attacked] at (2.3,-16.4) {}; \node[right=1.5mm] at (2.3,-16.4) {Attacked destinations};
\node[observed] at (19.5,-16.4) {}; \node[right=1.2mm] at (19.5,-16.4) {Attacked destinations with destination-specific telemetry};
\end{tikzpicture}%
}
\caption{Visibility mismatch in carpet bombing detection. Prefix-level traffic may cover the entire attacked surface while destination-specific telemetry is available for only a subset of attacked destinations.}
\label{fig:visibility_mismatch}
\end{figure}
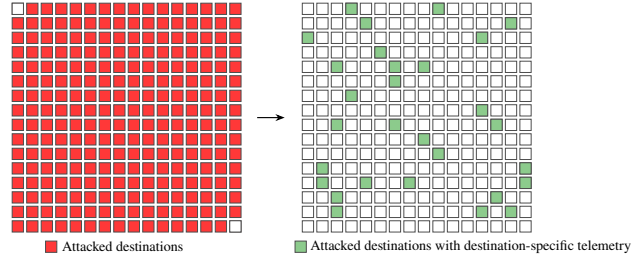

Figure~\ref{fig:visibility_mismatch} illustrates this mismatch. The targeted prefix may experience traffic across a broad attack surface, while only a subset of attacked destinations contributes destination-specific telemetry. Prefix-level measurements still reveal that abnormal traffic is reaching the network, but the strength of host-level evidence depends on the attack/monitor intersection.

This observation motivates the need for an attack-conditioned visibility metric. Rather than characterizing only how much of the prefix is active or instrumented, such a metric should quantify the fraction of the actual attack surface that is observable to the defender. The Prefix Visibility Ratio introduced in the next section provides this characterization.

\section{Prefix Visibility Ratio (PVR)}
\label{sec:pvr}

\subsection{Definition}

We define the time-dependent \emph{Prefix Visibility Ratio} (PVR) as
\begin{equation}
PVR_t=\frac{|A_t\cap O_t|}{|A_t|}, \qquad |A_t|>0,
\label{eq:pvr}
\end{equation}
where $A_t$ denotes the set of destinations receiving attack traffic during
observation window $t$, and $O_t$ denotes the set of destinations for which
the telemetry required by the host-level evidence plane is available to the defender. Router-side per-destination ingress measurements alone do not imply membership in $O_t$. The intersection
$A_t\cap O_t$ therefore represents the portion of the attacked surface that is
directly observable through host- or destination-specific evidence.

$PVR_t$ takes values in the interval $[0,1]$. At one extreme,
$PVR_t=1$ indicates that every attacked destination exposes
destination-specific evidence. At the other extreme, $PVR_t=0$ indicates that
none of the attacked destinations is directly observable at the destination
level. Intermediate values quantify partial attack visibility. For example,
$PVR_t=0.1$ means that only one tenth of the destinations receiving attack
traffic contribute destination-specific evidence to the detector, even though
prefix-level measurements may still observe the aggregate traffic delivered to
the network.

Equivalently, if one destination is selected uniformly from the attacked set
$A_t$, $PVR_t$ can be interpreted as the probability that this attacked
destination belongs to the defender's observable set $O_t$. This interpretation
highlights an important property of the metric: PVR is conditioned on the
attacker-selected destination set rather than on the prefix as a whole. It
therefore characterizes the visibility of a particular attack under a
particular monitoring deployment.

PVR is defined independently of attack intensity. Traffic volume, packet rate,
attack duration, and source population may affect the difficulty or impact of
an attack, but they do not directly enter Eq.~\eqref{eq:pvr}. Two attacks can
therefore generate the same aggregate rate toward the same prefix while
exhibiting substantially different PVR values if their destination sets overlap
differently with $O_t$. Conversely, two attacks with different traffic
intensities may have the same PVR if the same fraction of their attacked
destinations is observable.

The time index is also significant. Even when the monitoring deployment remains
unchanged, $PVR_t$ can vary across observation windows as the attacker changes
or rotates its target destinations. Likewise, changes in host availability or
telemetry collection can modify $O_t$ and therefore alter PVR without any
change in attack volume. PVR should consequently be viewed as a dynamic
property of the interaction between the attack surface and the monitoring
infrastructure, rather than as a static characteristic of the network.

The distinction among address utilization, nominal monitoring coverage, and
attack visibility can be summarized as
\begin{equation}
AUR_t=\frac{|U_t|}{|P|},\qquad
MC_t=\frac{|O_t|}{|U_t|},\qquad
PVR_t=\frac{|A_t\cap O_t|}{|A_t|}.
\label{eq:three_ratios}
\end{equation}
These quantities describe three different aspects of the deployment.
$AUR_t$ is \emph{prefix-population dependent}: it measures how much of the
usable address space is active. $MC_t$ is \emph{deployment dependent}: it
measures how much of the active population is instrumented. $PVR_t$ is
\emph{attack conditional}: it measures how much of the actual attacked
destination set is observable.

Importantly, neither $AUR_t$ nor $MC_t$, individually or jointly, determines
$PVR_t$. Consider two attacks against the same prefix while $U_t$ and $O_t$
remain unchanged. If the first attack primarily targets monitored destinations
and the second primarily targets unmonitored destinations, both scenarios have
identical AUR and monitoring coverage but potentially very different PVR
values. The missing information is precisely the overlap between $A_t$ and
$O_t$.

This property makes PVR particularly relevant to carpet bombing detection.
Such attacks deliberately distribute traffic over a destination set, making
the amount of destination-specific evidence dependent on which addresses are
selected. PVR provides a compact way to expose this otherwise implicit
observability condition and enables detector performance to be evaluated as a
function of the attack evidence actually available to the defender.

\subsection{Configured and Effective PVR in Finite Prefixes}
\label{sec:effective_pvr}

Finite address spaces make very small visibility ratios inherently discrete. In particular, when the attacked set contains only a small number of destinations, several nominal PVR values may map to the same integer number of observable attacked destinations. In our experiments, all ratio-to-count conversions therefore use conventional half-up rounding rather than Python's banker's rounding. For a positive configured ratio $q$ applied to a nonempty set of size $n$, the count used for active destinations, monitored destinations, and total attack extent is
\begin{equation}
N(q,n)=\min\!\left(n,\max\!\left(1,\left\lfloor nq+\tfrac{1}{2}\right\rfloor\right)\right).
\label{eq:half_up}
\end{equation}

For the requested monitored portion of the attacked set, the generator instead uses $M(q,n)=\min(n,\lfloor nq+\tfrac{1}{2}\rfloor)$ without a minimum-one constraint. This count is clipped to the feasible monitored/unmonitored allocation in the scenario, it can therefore be zero even when $q>0$. Attack-target selection subsequently chooses monitored and unmonitored destinations subject to those cardinalities. Because this process operates on finite sets, the realized overlap need not exactly match the nominal configured ratio. We therefore distinguish between the \emph{configured PVR}, used to define the experimental grid, and the \emph{effective PVR}, recomputed from the instantiated attack and observable sets:
\begin{equation}
PVR_t^{\mathrm{eff}}=\frac{|A_t\cap O_t|}{|A_t|}.
\end{equation}

The effective value is used whenever estimator accuracy is evaluated or whenever the realized attack/monitor overlap matters analytically. Consequently, nominal settings such as 1\%, 5\%, and 10\% may collapse to the same realized overlap when $|A_t|$ is small. We retain the configured labels to preserve a regular experimental grid, but we do not assume that these labels imply distinct realized cardinalities in every configuration. This distinction is especially important for sparse-PVR experiments on a /24 prefix, where a difference of a single destination can materially change the realized ratio.

\subsection{Visibility Regimes and Interpretation}

For presentation and discussion, we group PVR values into three descriptive regimes: full visibility ($PVR=1$), partial visibility ($0.1\leq PVR<1$), and sparse visibility ($PVR<0.1$). These categories provide a convenient way to summarize detector behavior across the observability range, but they should not be interpreted as universal operational thresholds.

In particular, we do not claim that $PVR\leq10\%$ is the dominant condition in real deployments, nor that 10\% represents a sharp transition in detection difficulty. The sparse regime is used primarily as a controlled stress condition in which destination-specific evidence becomes severely limited. Such situations are plausible when the attacked set includes inactive, unassigned, transient, or uninstrumented addresses, so that attack-conditioned overlap is substantially smaller than nominal monitoring coverage.

For the comparative summary, we use the separate label \emph{low-visibility configurations} for configured $PVR\leq10\%$, including the 10\% boundary. The regimes therefore serve as reporting conventions rather than assumptions about network operation. Their purpose is to expose how detector behavior changes as attacked-destination visibility deteriorates and to make low-observability failure modes explicit.

\section{Threat Model}
\label{sec:threat_model}

\subsection{Attacker Model}

We consider a distributed adversary capable of directing malicious traffic toward multiple destination addresses within an IPv4 prefix. Rather than concentrating traffic on a single victim, the attacker spreads the load across a destination set in order to threaten shared infrastructure such as access links, aggregation routers, or upstream connectivity while reducing the anomaly evidence visible at any individual host.

The attacker can vary the number of participating sources, the size and composition of the attacked destination set, the traffic intensity, and the duration of the attack. Traffic may be directed toward both active and inactive addresses, since packets sent to destinations that do not expose host telemetry can still consume forwarding and link capacity before being discarded or otherwise handled by the network.

The threat model does not grant the adversary knowledge of the defender's monitored-destination set $O_t$ or the ability to compromise, disable, or manipulate the monitoring infrastructure. To control visibility experimentally, however, the testbed generator uses $O_t$ to construct the requested mixture of monitored and unmonitored targets. This experimental conditioning is not an assumption of monitor-aware adversarial behavior, and target selection in the campaign is not statistically independent of the monitoring deployment.

A stronger monitor-aware adversary could attempt to infer which destinations are instrumented and deliberately bias or rotate its target set toward unmonitored addresses. Such an attacker would make $PVR_t$ itself an adversarially controlled quantity. This behavior is relevant to the broader visibility problem, but it is outside the scope of the present validated evaluation and is treated as a direction for future work rather than an assumption underlying the reported results.

\subsection{Defender Model}

The defender protects the target prefix using two complementary evidence planes. The first is a \emph{destination-specific plane}, which contains telemetry from the observable subset $O_t$ of active destinations. Depending on deployment capabilities, this telemetry may include received traffic rate, connection counts, source diversity, protocol distribution, and host resource-utilization indicators. These measurements provide detailed evidence about individual destinations but are available only where suitable instrumentation or telemetry exists.

The second is a \emph{network-level plane}, collected at a gateway, border router, or other aggregation point. It contains measurements such as aggregate traffic rate, destination dispersion, destination entropy, flow fan-out, source correlation, and, when available, per-destination ingress rates derived from network-side observation. These measurements characterize traffic reaching the protected prefix and remain available independently of host-side instrumentation.

The two planes are intentionally distinguished in the PVR definition. Network-side observation of packets addressed to a destination does not by itself place that destination in $O_t$. Membership in $O_t$ requires destination-specific evidence of the type used by the host-level component. Prefix-level measurements therefore provide complementary context rather than artificially increasing the attacked-destination visibility defined in Eq.~\eqref{eq:pvr}.

The defender is assumed to know which destinations currently provide destination-specific telemetry, but does not know the true attacked set $A_t$ or the true $PVR_t$ during operation. Both must be inferred from the available evidence.

\subsection{Design Objective}

The primary objective is binary prefix-level detection: determine whether the protected prefix is experiencing a carpet bombing attack while explicitly accounting for the fraction of the attacked surface represented in destination-specific telemetry. The objective is therefore distinct from victim localization, attack-source attribution, or mitigation, which are outside the scope of the present framework.

VANTA seeks to maintain reliable detection across changing observability conditions by exploiting the complementary properties of the two evidence planes. When destination-specific observations are sufficiently available, host-level evidence can contribute strongly to the decision. As visibility decreases, the framework should avoid interpreting missing host telemetry as benign evidence and instead place greater emphasis on prefix-level observations.

Crucially, VANTA does not assume that the true $PVR_t$ is available online. It estimates attack visibility from the telemetry itself and uses the resulting $\widehat{PVR}_t$ to adapt the fusion of host-level and prefix-level anomaly scores. The design objective is therefore not merely to detect attacks under partial monitoring, but to make the detector explicitly aware of the uncertainty introduced by incomplete attacked-destination visibility.

\section{VANTA: Visibility-Aware Network Threat Analysis}
\label{sec:vanta}

\subsection{System Overview}
\begin{figure}[t]
\centering
\resizebox{0.5\textwidth}{!}{%
\begin{tikzpicture}[
  font=\small,
  box/.style={draw,rounded corners=3pt,thick,align=center,minimum height=0.9cm,inner sep=6pt},
  input/.style={box,fill=green!10,minimum width=3.1cm},
  module/.style={box,fill=cyan!12,minimum width=3.6cm},
  detector/.style={box,fill=orange!12,minimum width=3.0cm},
  fusion/.style={box,fill=red!10,minimum width=4.0cm},
  arrow/.style={->,thick,>=Stealth}
]
\node[input] (host) at (-4.0,2.3) {Destination-specific\\host telemetry};
\node[input] (prefix) at (0,2.3) {Prefix/gateway\\telemetry};
\node[input] (inventory) at (4.0,2.3) {Monitored-destination\\inventory $O_t$};
\node[detector] (hd) at (-4.0,0.4) {Host-level detector\\$S_H(t)$};
\node[module] (ve) at (0,0.4) {Visibility estimator\\$\widehat{PVR}_t,\;c_t$};
\node[detector] (pd) at (4.0,0.4) {Prefix-level detector\\$S_P(t)$};
\node[fusion] (fu) at (0,-1.7) {Confidence-aware adaptive fusion\\$S(t)=w_HS_H+(1-w_H)S_P$};
\node[box,fill=red!55!black,text=white,minimum width=3.2cm] (alert) at (0,-3.4) {Carpet bombing alert};
\draw[arrow] (host) -- (hd); \draw[arrow] (host) -- (ve);
\draw[arrow] (prefix) -- (pd); \draw[arrow] (prefix) -- (ve);
\draw[arrow] (inventory) -- (ve);
\draw[arrow] (hd) -- (fu); \draw[arrow] (pd) -- (fu); \draw[arrow] (ve) -- (fu);
\draw[arrow] (fu) -- (alert);
\end{tikzpicture}%
}
\caption{VANTA architecture. Destination-specific and prefix-level telemetry feed independent anomaly detectors, while the visibility estimator derives $\widehat{PVR}_t$ and confidence $c_t$. These quantities control the adaptive fusion of host- and prefix-level evidence into the final carpet bombing decision.}
\label{fig:vanta_architecture}
\end{figure}
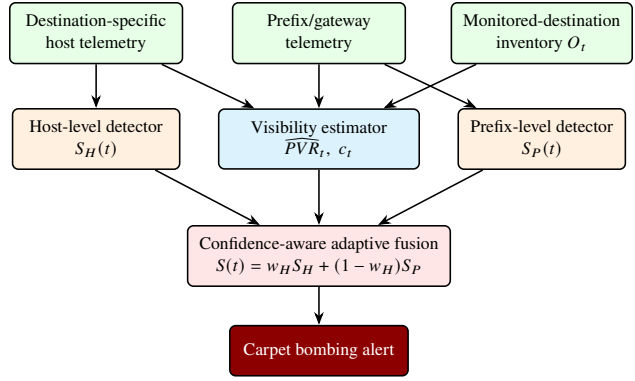

Figure~\ref{fig:vanta_architecture} presents the overall VANTA architecture. The framework separates destination-specific evidence from prefix-level network evidence and processes them through two complementary detection paths. Destination-specific telemetry from the monitored set $O_t$ feeds the host-level detector, while gateway or router measurements feed the prefix-level detector. The two planes are deliberately analyzed separately because their reliability changes differently as attacked-destination visibility decreases.

In parallel, the visibility estimator uses network-side telemetry together with the monitored-destination inventory to infer the fraction of the attacked set that is observable. When sufficiently detailed router-side measurements are available, it estimates the attacked destination set directly, otherwise, it falls back to destination-cardinality and host-evidence estimates. The estimator outputs both the online visibility estimate $\widehat{PVR}_t$ and a confidence value $c_t$.

These two quantities control the adaptive fusion stage. Rather than assigning fixed importance to host- and prefix-level evidence, VANTA adjusts their relative contribution according to the estimated visibility of the attack and the confidence placed in that estimate. This design prevents sparse host telemetry from being interpreted as evidence of benign behavior and allows prefix-level signals to dominate when destination-specific observations become unreliable.

\subsection{Online PVR Estimation}
\label{subsec:visibility_estimation}

VANTA estimates PVR online from observable telemetry and experimental ground truth is never supplied to the estimator. The validated implementation restricts attacked-set inference to active destinations, matching the campaign assumption $A_t\subseteq U_t$. The general definition in Eq.~\eqref{eq:pvr} also permits attacks outside $U_t$, but that case is not evaluated here. The estimation procedure has two goals: identify which destinations are likely to be under attack and determine what fraction of that estimated attacked set belongs to the known monitored-destination set $O_t$.

During the benign calibration phase, VANTA learns an address-specific ingress-rate baseline for each active destination using router-side per-destination measurements. For destination $j$, let $\widetilde r_j$ denote the benign median ingress rate, $MAD_j$ the corresponding median absolute deviation, and
$\sigma_j^{R}=1.4826\,MAD_j$ the robust standard-deviation estimate. The destination-specific anomaly threshold is
\begin{equation}
T_j=\max\left(
r_{\min},
\widetilde r_j+\gamma\max(\sigma_j^R,\sigma_{\min}),
r_j^{\max}+\sigma_{\min}
\right),
\label{eq:destination_threshold}
\end{equation}
where $r_j^{\max}$ is the largest rate observed during benign calibration. In the validated configuration, $\gamma=6$, $\sigma_{\min}=1000$ b/s, and $r_{\min}=1$ b/s. The final term acts as a training-maximum guard, ensuring that all benign calibration samples remain below the learned attack threshold.

When router-side per-destination rates are available, VANTA estimates the attacked destination set as
\begin{equation}
\widehat A_t=\{j\in U_t:r_{j,t}>T_j\}.
\label{eq:attacked_set_estimate}
\end{equation}
Because $O_t$ represents deployment knowledge rather than attack ground truth, the estimated number of observable attacked destinations is obtained directly from
$|\widehat A_t\cap O_t|$. The resulting raw visibility estimate is
\begin{equation}
\widehat{PVR}^{\,raw}_t=
\begin{cases}
\dfrac{|\widehat A_t\cap O_t|}{|\widehat A_t|},
& |\widehat A_t|>0,\\[1ex]
0, & \text{otherwise.}
\end{cases}
\label{eq:pvr_raw}
\end{equation}

To reduce sensitivity to short-lived fluctuations in the inferred attacked set, the raw estimate is temporally smoothed using an exponentially weighted moving average:
\begin{equation}
\widehat{PVR}_t=
\beta\widehat{PVR}^{\,raw}_t+
(1-\beta)\widehat{PVR}_{t-1},
\label{eq:pvr_ewma}
\end{equation}
with $\beta=0.35$ in the validated configuration.

The direct set-based path is preferred whenever per-destination router measurements are available. When they are not, VANTA falls back to an approximate attacked-set cardinality inferred from prefix-level destination dispersion, while suspicious monitored hosts provide the corresponding observable-attacked estimate. This fallback allows the estimator to remain operational with coarser network telemetry, although with lower confidence than the direct per-destination path. The implementation records which denominator-estimation path is used.

Host evidence used by the fallback combines normalized received rate, source diversity, UDP fraction, and resource utilization with respective weights $0.45$, $0.30$, $0.20$, and $0.05$. A host is treated as suspicious for visibility estimation when this score reaches $0.55$. Independently, prefix attack likelihood combines aggregate traffic rate, destination dispersion, destination entropy, flow fan-out, and source correlation with weights $0.30$, $0.25$, $0.15$, $0.15$, and $0.15$, respectively, and the associated attack-evidence threshold is $0.40$.

Because not all visibility estimates are equally reliable, the estimator also produces an explicit confidence score
\begin{equation}
c_t=\mathrm{clip}_{[0,1]}
\!\left(
0.45c_A+
0.25q_P+
0.20\rho_O+
0.10\overline q_H
\right),
\label{eq:visibility_confidence}
\end{equation}
where $c_A$ represents the reliability of the attacked-set estimate, $q_P$ the prefix-level attack likelihood, $\rho_O$ the reporting fraction relative to the estimated attacked-set size, and $\overline q_H$ the mean host-evidence score. Positive confidence values are lower-bounded by $0.05$.

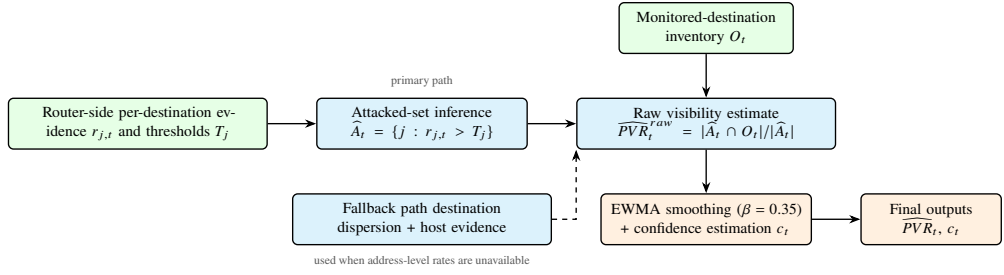
\begin{figure}[t]
\centering
\resizebox{0.8\textwidth}{!}{%
\begin{tikzpicture}[
 font=\small,
 node distance=9mm and 10mm,
 box/.style={
   draw,
   rounded corners=3pt,
   thick,
   align=center,
   minimum height=1.05cm,
   text width=3.25cm,
   inner sep=5pt
 },
 input/.style={box,fill=green!10},
 proc/.style={box,fill=cyan!12},
 output/.style={box,fill=orange!12},
 arrow/.style={->,thick,>=Stealth},
 fallbackarrow/.style={->,thick,dashed,>=Stealth}
]

\node[input, text width=5cm] (router) {Router-side per-destination evidence $r_{j,t}$ and thresholds $T_j$};

\node[proc, right=of router, text width=4cm] (aset)
{Attacked-set inference\\
$\widehat A_t=\{j:r_{j,t}>T_j\}$};

\node[proc, right=of aset, text width=5cm] (ratio)
{Raw visibility estimate\\
$\widehat{PVR}^{\,raw}_t=
|\widehat A_t\cap O_t|/|\widehat A_t|$};

\node[input, above=of ratio] (inventory)
{Monitored-destination inventory $O_t$};

\node[proc, below=of aset, text width=5cm] (fallback)
{Fallback path destination dispersion + host evidence};

\node[output, below=of ratio, text width=4cm] (smooth)
{EWMA smoothing ($\beta=0.35$) + confidence estimation $c_t$};

\node[output, right=of smooth, text width=2.55cm] (out)
{Final outputs\\$\widehat{PVR}_t$, $c_t$};

\draw[arrow] (router) -- (aset);
\draw[arrow] (aset) -- (ratio);
\draw[arrow] (inventory) -- (ratio);
\draw[fallbackarrow] (fallback.east) -| (ratio.south west);
\draw[arrow] (ratio) -- (smooth);
\draw[arrow] (smooth) -- (out);

\node[font=\scriptsize, text=black!65, above=1mm of aset]
{primary path};

\node[font=\scriptsize, text=black!65, below=1mm of fallback]
{used when address-level rates are unavailable};

\end{tikzpicture}
}
\caption{Online PVR-estimation pipeline. The primary path infers the attacked destination set from router-side per-destination rate anomalies and intersects it with the monitored-destination inventory $O_t$. When address-level network measurements are unavailable, destination dispersion and host evidence provide a fallback estimate. The resulting raw PVR is temporally smoothed and accompanied by an explicit confidence score.}
\label{fig:pvr_estimation}
\end{figure}

The confidence variable serves an important role in the subsequent fusion stage: two identical values of $\widehat{PVR}_t$ need not have the same influence when one is supported by direct per-destination measurements and the other is obtained primarily through fallback inference. Figure~\ref{fig:pvr_estimation} summarizes the complete estimation pipeline.

\subsection{Host- and Prefix-Level Anomaly Scores}
\label{subsec:multi_level_detection}

VANTA maintains separate anomaly scores for destination-specific and prefix-level evidence. This separation is essential because the two evidence sources respond differently to partial observability: host-level evidence weakens as fewer attacked destinations are represented in $O_t$, whereas aggregate prefix evidence remains available independently of destination-specific instrumentation.

For each monitored host $i$, the host detector operates on
\begin{equation}
\mathbf{x}_{i,t}^{H}
=
[r_{i,t},c_{i,t},s_{i,t},p_{i,t},u_{i,t}],
\end{equation}
where the components represent received traffic rate, connection rate, source diversity, protocol-related traffic composition, and resource utilization. These features capture complementary indicators of abnormal host behavior, ranging from increased traffic intensity to changes in source population and protocol composition.

At the network level, the prefix detector uses
\begin{equation}
\mathbf{x}_{t}^{P}
=
[R_t,D_t,E_t,F_t,C_t],
\end{equation}
where $R_t$ is aggregate traffic rate, $D_t$ destination dispersion, $E_t$ destination entropy, $F_t$ flow fan-out, and $C_t$ source correlation. Unlike the host vector, these features characterize how traffic is distributed across the protected prefix and are therefore particularly useful when destination-specific visibility is limited.

Both detectors are calibrated on benign traffic. For each feature $k$, VANTA learns a benign mean $\mu_k$ and standard deviation $\sigma_k$ and computes a one-sided standardized anomaly score
\begin{equation}
s_k(x)=
\min\left(
1,
\max\left(
0,
\frac{x_k-\mu_k}{\lambda\sigma_k}
\right)
\right),
\label{eq:one_sided_score}
\end{equation}
with $\lambda=3$. The one-sided formulation reflects the attack model considered here: the relevant indicators are expected to increase under carpet bombing, while values below the benign mean provide no positive attack evidence.

The detector-level score is the arithmetic mean of the available feature scores. For the host plane, individual host scores are subsequently averaged over the monitored destinations that report during the current observation window, yielding $S_H(t)$. The prefix features are combined analogously to obtain $S_P(t)$.

Some network features require evidence from multiple destinations. In particular, flow fan-out and source correlation are excluded when fewer than two reporting destinations are available because their cross-destination interpretation is not meaningful in that regime. This feature-availability rule prevents unsupported measurements from artificially influencing the prefix score.

The resulting $S_H(t)$ and $S_P(t)$ intentionally remain independent at this stage. Their relative importance is determined only in the adaptive-fusion stage, where $\widehat{PVR}_t$ and estimator confidence $c_t$ determine how much trust should be placed in destination-specific versus prefix-level evidence.

\subsection{Confidence-Aware Adaptive Fusion}
\label{subsec:adaptive_fusion}

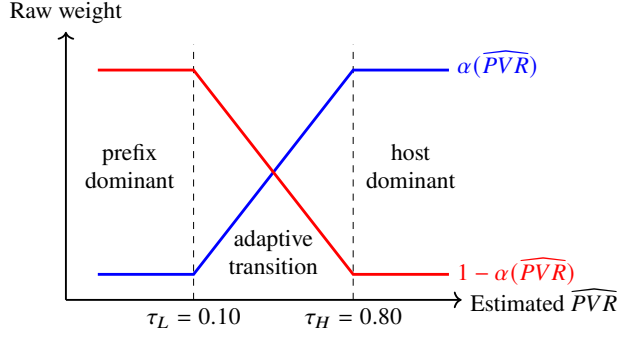
\begin{figure}[t]
\centering
\resizebox{0.50\textwidth}{!}{%
\begin{tikzpicture}[scale=0.95]
\draw[->,thick] (0,0) -- (6.2,0) node[right] {Estimated $\widehat{PVR}$};
\draw[->,thick] (0,0) -- (0,4.2) node[above] {Raw weight};
\draw[dashed] (2,0) -- (2,4); \draw[dashed] (4.5,0) -- (4.5,4);
\node[below] at (2,0) {$\tau_L=0.10$}; \node[below] at (4.5,0) {$\tau_H=0.80$};
\draw[very thick,blue] (0.5,0.4)--(2,0.4)--(4.5,3.6)--(6,3.6);
\draw[very thick,red] (0.5,3.6)--(2,3.6)--(4.5,0.4)--(6,0.4);
\node[blue,right] at (6.0,3.7) {$\alpha(\widehat{PVR})$};
\node[red,right] at (6.0,0.45) {$1-\alpha(\widehat{PVR})$};
\node[align=center] at (1.0,2.1) {prefix\\dominant};
\node[align=center] at (3.25,0.75) {adaptive\\transition};
\node[align=center] at (5.4,2.1) {host\\dominant};
\end{tikzpicture}%
}
\caption{PVR-dependent fusion weights before confidence scaling. For $\widehat{PVR}\leq\tau_L$, host evidence receives no raw weight; between $\tau_L$ and $\tau_H$, its contribution increases linearly; and above $\tau_H$, the raw host weight reaches one. The effective host contribution is further scaled by estimator confidence, $w_H=\alpha(\widehat{PVR})c_t$.}
\label{fig:adaptive_weighting}
\end{figure}

The final VANTA decision combines host-level and prefix-level anomaly evidence according to the estimated visibility of the attack. The first stage derives a raw host weight from the online PVR estimate:
\begin{equation}
\alpha(\widehat{PVR}_t)=
\min\!\left(
1,
\max\!\left(
0,
\frac{\widehat{PVR}_t-\tau_L}{\tau_H-\tau_L}
\right)
\right).
\label{eq:alpha}
\end{equation}
In the validated configuration, $\tau_L=0.10$ and $\tau_H=0.80$. Thus,
host evidence receives no raw weight when estimated visibility is at or below
$\tau_L$, becomes progressively more influential in the transition interval
$(\tau_L,\tau_H)$, and reaches full raw weight when
$\widehat{PVR}_t\geq\tau_H$. Figure~\ref{fig:adaptive_weighting} illustrates
this piecewise-linear behavior.

PVR alone, however, does not indicate how reliable the visibility estimate is.
VANTA therefore modulates the raw weight using estimator confidence $c_t$.
With the validated linear confidence scaling and no additional confidence floor
inside the fusion module, the effective weights are
\begin{equation}
w_H(t)=\alpha(\widehat{PVR}_t)c_t,
\qquad
w_P(t)=1-w_H(t).
\label{eq:effective_alpha}
\end{equation}
The fused anomaly score is then
\begin{equation}
S(t)=w_H(t)S_H(t)+w_P(t)S_P(t).
\label{eq:fusion_score}
\end{equation}

This construction has two useful properties. First, low estimated PVR
automatically shifts the decision toward prefix-level evidence, whose
availability does not depend on destination-specific instrumentation. Second,
even when $\widehat{PVR}_t$ is high, uncertain visibility estimation prevents
the detector from placing excessive trust in host-level evidence. Conversely,
host evidence becomes dominant only when both estimated visibility and
estimator confidence are sufficiently high.

The general fused-score detection threshold is $\theta=0.60$. If
$w_H(t)=0$ and the estimator has inferred at least one attacked destination,
the prefix-only score threshold is $0.40$. When no attacked destination is
inferred, the general $0.60$ threshold remains in force. In addition, the
implementation retains two corroboration gates. Host support can trigger an
alert when $S_H\geq0.60$, at least one monitored destination is estimated to
be attacked, and at least one host is flagged as suspicious. This gate applies
in the estimator's \texttt{low\_visibility} and \texttt{hybrid} modes,
corresponding respectively to $0<\widehat{PVR}_t<0.25$ and
$\widehat{PVR}_t\geq0.25$, provided the inferred attacked set is nonempty.
Prefix support can trigger an alert when $S_P\geq1/3$ in the estimator's
\texttt{prefix\_only} mode, which requires a nonempty inferred attacked set
and $\widehat{PVR}_t=0$. If no attacked destination is inferred, the mode is
\texttt{no\_attack\_evidence} and neither corroboration gate applies. An alert
is raised if the applicable fused-score threshold or either eligible gate is
satisfied. These gates do not modify $\widehat{PVR}_t$, $c_t$, or the
underlying anomaly scores, they affect only the final decision logic.

We report all fusion and decision thresholds explicitly because they determine
how estimated observability is translated into operational detector behavior.
In particular, $\tau_L$ and $\tau_H$ control the transition between
prefix-dominant and host-dominant evidence, whereas $\theta$ and the
corroboration gates determine when the resulting evidence is sufficient to
raise an alert.

\subsection{Key Implementation Parameters}

For reproducibility, Table~\ref{tab:key_parameters} summarizes the principal
parameters fixed in the validated VANTA implementation. These values
control anomaly scoring, PVR estimation, adaptive fusion, and the final
decision logic.

The parameters fall into three main groups. The first controls host- and
prefix-level anomaly detection, including the statistical sensitivity
$\lambda$ and the detector thresholds. The second governs online PVR
estimation, including EWMA smoothing, evidence thresholds, and robust
destination-rate calibration. The third controls confidence-aware fusion and
the transition between prefix-dominant and host-dominant operation. All values
reported below were fixed for the validated experimental campaign rather than
tuned separately for individual PVR, AUR, or attack configurations.

\begin{table}[H]
\centering
\caption{Key VANTA parameters used throughout the validated campaign.}
\label{tab:key_parameters}
\small
\begin{tabular}{@{}ll@{}}
\toprule
Parameter & Value \\
\midrule
Statistical sensitivity $\lambda$ & 3.0 \\
Host detector threshold & 0.60 \\
Prefix detector threshold & 0.40 \\
Fusion $\tau_L,\tau_H$ & 0.10, 0.80 \\
General fusion threshold $\theta$ & 0.60 \\
Prefix-only decision threshold & 0.40 \\
PVR-estimator EWMA $\beta$ & 0.35 \\
Estimator low-visibility mode boundary & 0.25 \\
PVR-estimator host evidence threshold & 0.55 \\
PVR-estimator prefix evidence threshold & 0.40 \\
Destination-rate robust scale $\gamma$ & 6.0 \\
Destination baseline std. floor & 1000 b/s \\
Host/prefix corroboration thresholds & 0.60 / $1/3$ \\
\bottomrule
\end{tabular}
\end{table}

These parameters define the operating point used for all results reported in
this paper. We expose them explicitly to make the decision process
reproducible and to distinguish fixed implementation choices from experimental
variables such as PVR, AUR, attack extent, and source population.

\section{Experimental Methodology}
\label{sec:methodology}

\subsection{Testbed and Traffic Model}

We implement VANTA in Mininet using a /24 target prefix containing 253 usable destination addresses (excluding the network and broadcast addresses and one host address reserved for the gateway) and a single aggregation router through which all benign and attack traffic enters the protected network. Each instantiated topology includes 50 legitimate clients generating background traffic and 100 distributed attack sources capable of directing malicious traffic toward multiple destinations within the prefix. These population sizes are kept fixed across the validated campaign so that observability and attack-distribution effects can be studied under a consistent traffic-generation environment.

Prefix occupancy is controlled through the Address Utilization Ratio,
$AUR\in\{0.1,0.2,0.5,1.0\}$, corresponding to 25, 51, 127, and 253 active destinations, respectively, after the half-up integer conversion defined in Eq.~\eqref{eq:half_up}. The remaining addresses are treated as inactive for that topology. Although the general threat model permits traffic toward inactive addresses, the validated campaign selects attack targets only from active destinations, so $A_t\subseteq U_t$. Targets need not belong to the monitored subset $O_t$.

Destination-specific telemetry is available only for the monitored subset of active hosts. These observations provide the host-level evidence used by VANTA, including traffic and host-specific statistics. In contrast, prefix-level measurements are collected at the aggregation router and remain available independently of which individual destinations are monitored. This arrangement reproduces the two evidence planes assumed by the defender model: partial destination-specific visibility combined with continuous network-level observation.

The testbed controls address utilization, monitored-destination availability, and attack-target selection explicitly. In the present campaign, the configured PVR parameter also determines the monitored fraction of active destinations. Monitoring coverage and attack visibility are therefore coupled by design, rather than varied independently. Identical prefix occupancy can nevertheless produce different attack visibility as the attacked/monitored overlap changes.

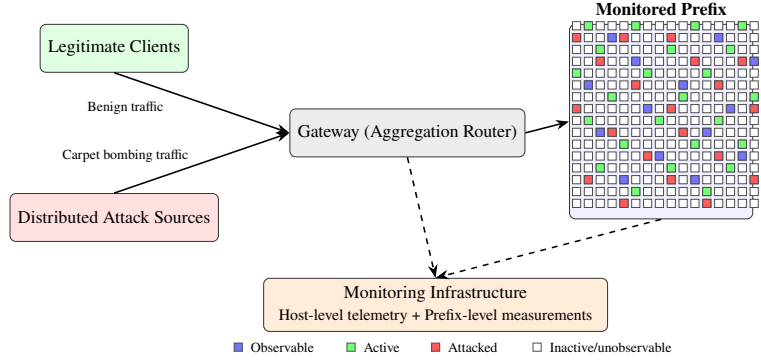
\begin{figure}
\centering
\resizebox{0.6\textwidth}{!}{%
\begin{tikzpicture}[
    font=\small,
    block/.style={
        rectangle,
        rounded corners=3pt,
        draw=black!80,
        line width=0.45pt,
        align=center,
        inner sep=4pt,
        minimum height=0.85cm
    },
    source/.style={
        block,
        fill=green!12,
        minimum width=2.6cm
    },
    attack/.style={
        block,
        fill=red!12,
        minimum width=2.6cm
    },
    router/.style={
        block,
        fill=gray!15,
        minimum width=2.6cm
    },
    prefixbox/.style={
    rectangle,
    rounded corners=3pt,
    draw=black!80,
    line width=0.45pt,
    fill=blue!6,
    align=center,
    inner sep=5pt
    },
    monitor/.style={
        block,
        fill=orange!15,
        minimum width=6.1cm
    },
    cell/.style={
        rectangle,
        draw=black!70,
        line width=0.18pt,
        minimum width=1.45mm,
        minimum height=1.45mm,
        inner sep=0pt,
        outer sep=0pt
    },
    active/.style={cell, fill=green!55},
    observable/.style={cell, fill=blue!55},
    attacked/.style={cell, fill=red!65},
    inactive/.style={cell, fill=white},
    arrow/.style={->, thick, >=Stealth},
    dashedarrow/.style={->, thick, dashed, >=Stealth}
]

\node[source] (clients) at (-5.2,1.8)
{Legitimate Clients};

\node[attack] (attackers) at (-5.2,-1.2)
{Distributed Attack Sources};

\node[router] (gateway) at (0,0.3)
{Gateway (Aggregation Router)};

\node[prefixbox, minimum width=3.25cm, minimum height=3.45cm] (prefixbox) at (4.5,0.5) {};
\node[font=\small\bfseries] at (4.5,2.5) {Monitored Prefix};

\begin{scope}[xshift=3.0cm,yshift=2.2cm,x=2.1mm,y=2.1mm]

\foreach \x in {0,...,15}{
  \foreach \y in {0,...,15}{
    \node[inactive] at (\x,-\y) {};
  }
}

\foreach \x/\y in {
1/0, 5/0, 10/0, 14/0,
2/2, 8/2, 13/2,
0/4, 6/4, 11/4,
3/6, 9/6, 15/6,
1/8, 7/8, 12/8,
4/10, 10/10, 14/10,
2/12, 8/12, 13/12,
5/14, 11/14
}{
  \node[active] at (\x,-\y) {};
}

\foreach \x/\y in {
3/1, 12/1,
5/3, 15/3,
1/5, 9/5,
6/7, 13/7,
2/9, 11/9,
7/11, 14/11,
4/13, 10/13
}{
  \node[observable] at (\x,-\y) {};
}

\foreach \x/\y in {
0/1, 4/1, 8/1,
2/3, 10/3, 14/3,
5/5, 12/5,
0/7, 8/7, 15/7,
3/9, 9/9,
6/11, 12/11,
1/13, 8/13, 15/13,
4/15, 11/15
}{
  \node[attacked] at (\x,-\y) {};
}

\end{scope}

\node[monitor] (monitor) at (0.5,-2.75)
{Monitoring Infrastructure\\
\footnotesize Host-level telemetry + Prefix-level measurements};

\draw[arrow] (clients.south) -- (gateway.west);
\draw[arrow] (attackers.north) -- (gateway.west);

\draw[arrow] (gateway.east) -- (prefixbox.west);

\draw[dashedarrow] (gateway.south) -- (monitor.north);
\draw[dashedarrow] (prefixbox.south) -- (monitor.north);

\node[font=\scriptsize, align=center] at (-5.0,0.75) {Benign traffic};
\node[font=\scriptsize, align=center] at (-5.0,-0.1) {Carpet bombing traffic};

\node[observable] at (-3.0,-3.5) {};
\node[right=1.0mm] at (-3.0,-3.5) {\scriptsize Observable};

\node[active] at (-1.0,-3.5) {};
\node[right=1.0mm] at (-1.0,-3.5) {\scriptsize Active};

\node[attacked] at (0.5,-3.5) {};
\node[right=1.0mm] at (0.5,-3.5) {\scriptsize Attacked};

\node[inactive] at (2.3,-3.5) {};
\node[right=1.0mm] at (2.3,-3.5) {\scriptsize Inactive/unobservable};

\end{tikzpicture}%
}
\caption{Mininet-based emulation testbed. Legitimate clients and distributed attack sources send traffic through a common aggregation router toward the target /24 prefix. Destination-specific telemetry is collected only from the monitored host subset, while prefix-level measurements remain available at the gateway, enabling controlled evaluation under different AUR and PVR conditions.}
\label{fig:testbed_architecture}
\end{figure}

Figure~\ref{fig:testbed_architecture} summarizes the resulting topology. Legitimate clients and distributed attack sources send traffic through the common gateway toward the protected prefix, while the monitoring infrastructure collects both router-side measurements and telemetry from the observable host subset.

\subsection{Visibility and Attack Configurations}

For each address-utilization level, we evaluate six configured visibility
conditions:
\[
PVR \in \{0.01,\,0.05,\,0.10,\,0.20,\,0.50,\,1.00\}.
\]
Carpet bombing attacks are instantiated with destination extents of
25\%, 50\%, or 100\% of the active address set.
Combining the four AUR values, six configured PVR values, and three attack
extents yields 72 distinct AUR--PVR--attack-extent configurations per
repetition. Each configuration is executed 30 times using deterministic but
distinct random seeds, resulting in 24 topology instances per repetition,
720 instantiated topologies overall, and 2,160 completed experimental runs.

For each configured PVR value $q$, the generator first sets
$|O_t|=N(q,|U_t|)$ using Eq.~\eqref{eq:half_up}.
The nominal monitoring coverage is therefore
$MC_t=N(q,|U_t|)/|U_t|$. For attack extent $f$, it sets
$|A_t|=N(f,|U_t|)$.
It then requests $m=\lfloor q|A_t|+\tfrac{1}{2}\rfloor$ monitored attack targets, clipped to
the feasible interval
$[\max(0,|A_t|-|U_t\setminus O_t|),\min(|A_t|,|O_t|)]$.
Monitored and unmonitored targets are sampled within their respective sets.
Unlike active, monitored, and total target counts, the monitored attack-target
count is not forced to be positive. Thus a positive configured PVR can yield
$PVR_t^{\mathrm{eff}}=0$. The published campaign contains 180 such runs.

The realized attack set and its intersection with the observable set are
recorded explicitly for every run. Consequently, configured PVR is a target
experimental condition, not an exact guarantee on the realized ratio. The
campaign couples monitoring coverage and target overlap through $q$. It does
not separately sweep $MC_t$ at fixed effective PVR.

This distinction is particularly important in the sparse-visibility regime.
For small attacked sets, different nominal values such as 1\%, 5\%, and 10\%
may map to the same number of observable attacked destinations after integer
conversion, and can therefore produce identical effective overlap. For this
reason, configured PVR values are retained as labels for the experimental grid,
whereas all visibility-estimation accuracy calculations use the realized
effective PVR,
$PVR^{\mathrm{eff}}_t=|A_t\cap O_t|/|A_t|$.

\subsection{Calibration, Evaluation Windows, and Metrics}

Each primary VANTA run is divided into three successive phases: 20 benign
calibration windows, five benign evaluation windows, and ten attack windows.
The calibration phase is used exclusively to learn the benign reference
statistics required by the host detector, prefix detector, and
destination-specific PVR estimator. These windows are not included in the
reported detection metrics.

The subsequent benign windows evaluate false-alarm behavior under traffic not
used for calibration, while the attack windows measure the detector response
after carpet bombing begins. A run in which no alert is raised during the
attack period is retained as a false negative. Such runs are excluded only
from the conditional detection-delay statistic, since no finite detection time
exists, but they remain fully accounted for in Recall, F1-score, and the other
classification metrics.

We report Precision, Recall, F1-score, False Positive Rate (FPR), Accuracy,
detection delay, and absolute PVR-estimation error. Delay is reported only
for successfully detected attacks using the campaign convention
$d=\delta+k\Delta$, where $k\in\{1,\ldots,10\}$ is the first detected
attack-window index and $\Delta=1$~s is the configured window interval. The
published records include a pre-evaluation offset of $\delta=3$~s: first-window
detection is recorded as 4~s and last-window detection as 13~s. This is a
window-index delay including the offset, rather than a measurement of detector
processing latency. The offset is not counted as an evaluated attack window.

For PVR accuracy, the implementation first averages the online estimates over
the ten attack windows of each run $r$, then computes the absolute deviation
from that run's constant realized effective visibility:
\begin{equation}
e_{PVR}^{(r)}=
\left|
\frac{1}{T}\sum_{t=1}^{T}\widehat{PVR}_{r,t}-PVR_r^{\mathrm{eff}}
\right|,\qquad T=10.
\label{eq:pvr_error}
\end{equation}
The reported mean, median, and error fractions summarize these per-run values.
They are not the mean or median of window-level absolute errors. Temporal
over- and under-estimation may cancel within a run.

Classification metrics are derived from pooled confusion matrices across the
relevant runs rather than by averaging independently computed per-run ratios.
This aggregation preserves the underlying numbers of true positives, true
negatives, false positives, and false negatives and avoids giving equal weight
to runs with different numbers of evaluated decisions.

\subsection{Task-Adapted Reference Implementations}

We evaluate three lightweight and feature-level reference implementations and
append the suffix ``-S'' to their names to distinguish them explicitly from the
original end-to-end systems. The goal is not to reproduce every architectural
detail or original task objective, but to preserve representative feature
families and decision structures within a common binary prefix-level detection
setting.

NetRadar-S retains three core forms of evidence from NetRadar: traffic-derived
features, server-assisted features augmented with an availability mask, and
permutation-invariant cross-destination context. The original PointNet-based
processing pipeline is replaced by a Random Forest so that the reference model
can be evaluated consistently within our binary detection framework. The
resulting implementation should therefore be interpreted as a task-adapted
reference inspired by NetRadar rather than as an exact reproduction of the
original system.

FlowLens-S preserves the main statistical representation used by FlowLens,
including normalized packet-size and inter-arrival histograms, preliminary
feature-importance ranking, selection of the six most informative features, and
Random-Forest classification~\cite{flowlens}. Whisper-S retains the original
160-packet size--protocol--inter-arrival representation, partitions it into five
32-point FFT blocks, and uses 16 spectral coefficients. Classification follows
the supervised Random-Forest adaptation used in the NetRadar
benchmark~\cite{netradar,whisper}.

These three references are used for a specific experimental purpose: to examine
how reduced destination observability affects methods based on different
traffic representations when all are evaluated on the same binary question,
namely whether the protected prefix is experiencing a carpet bombing attack.
They are therefore intended to provide controlled points of comparison under
matched observability conditions, not to establish end-to-end superiority over
the original systems.

This distinction is particularly important for NetRadar. Its original objective
includes identifying or localizing attacked servers, whereas VANTA evaluates
prefix-level attack presence. A failure to localize a specific victim does not
imply that an ongoing attack is undetectable from gateway-side aggregate
evidence. Accordingly, our NetRadar-S results should be interpreted only as a
comparison of binary detection behavior under controlled destination
observability, not as a reproduction of NetRadar's victim-localization
performance.

\subsection{Observability Modes and Data Splits}

We evaluate the reference methods under two observability modes. In the
\emph{full-visibility} mode, each baseline receives packet- or flow-level
features for every active destination. These results are reported only as
optimistic upper bounds because they assume destination-specific evidence that
may not be available in realistic partial-monitoring conditions.

In the \emph{matched-observability} mode, baseline training, validation, and
testing are restricted to destinations marked as monitored in the corresponding
scenario. The reference methods therefore operate under the same
destination-specific coverage constraint used to define PVR. Per-destination
prediction probabilities are converted into a prefix-level score by averaging
the highest 10\% of the available destination scores, providing a common binary
decision statistic across baselines. Detection thresholds are selected using
validation data only and are then fixed for the held-out test repetitions.

The comparative campaign exports five benign and ten attack windows from each
run, yielding 32,400 one-second windows in total. Repetitions 1--10 are used
for baseline training, repetitions 11--15 for threshold validation, and
repetitions 16--30 for final testing. To limit class imbalance and training-set
size, each baseline retains at most 15,000 positive and 15,000 negative
training examples and up to 6,000 examples per class for validation.

VANTA's comparative results are taken from the independently validated primary
campaign, restricted to the same held-out repetition indices and
configuration/seed schedule. The comparison is therefore seed-aligned at the
scenario level, but it is not packet-for-packet paired across implementations.
For each held-out repetition, comparative figures first pool all AUR values and
attack extents associated with a given observability condition. We then report
the mean across the 15 test repetitions together with the corresponding
two-sided Student-$t$ 95\% confidence interval.

\subsection{Scope of the Current Evaluation}

The present campaign is designed primarily to characterize the effect of
attacked-destination observability under controlled conditions. It is not
intended to estimate the prevalence of a particular PVR regime in operational
networks, nor to reproduce ISP-scale traffic, addressing, or deployment
conditions.

The /24 Mininet environment and synthetic traffic model are deliberately used
to control AUR, attack extent, monitoring availability, and the overlap between
attacked and observable destinations explicitly. Monitoring coverage and
attack overlap are coupled as described in Section~\ref{sec:methodology}. The campaign characterizes visibility under that construction, rather than
isolating it from every monitoring-deployment effect. Consequently, sparse-PVR configurations should be
interpreted as observability stress cases rather than as claims about typical
production deployments.

Several questions therefore remain outside the scope of the current validated
campaign. These include evaluation on real traffic traces and larger prefixes,
adaptive attackers that rotate targets according to inferred monitoring
coverage, exact end-to-end reproductions of the reference systems, component
ablations, oracle-PVR comparisons, and sensitivity analysis for fusion and
decision parameters. These extensions form the natural next stage of the
evaluation, but are not assumed by the results reported in this paper.

\section{Experimental Results}
\label{sec:vanta_evaluation}

\subsection{Intrinsic VANTA Performance}

Across the complete validated campaign, VANTA processes 21,600 attack windows
and 10,800 benign evaluation windows. The resulting pooled confusion matrix
contains 20,187 true positives, 10,634 true negatives, 166 false positives,
and 1,413 false negatives. This corresponds to a precision of 99.18\%, a
recall of 93.46\%, a global F1-score of 96.24\%, a false-positive rate (FPR)
of 1.54\%, and a false-negative rate (FNR) of 6.54\%. These results indicate
that, when all experimental conditions are pooled, VANTA maintains high
detection sensitivity while generating relatively few false alarms.

The mean F1-score computed independently at the run level is 94.76\%, slightly
below the pooled value of 96.24\%. The difference results from the aggregation
method: pooled F1 weights decisions according to the total number of windows,
whereas the per-run mean gives equal importance to every run. Consequently,
runs in which the detector completely misses an attack receive the same
macro-level weight as otherwise well-performing runs and have a stronger
effect on the mean per-run score. Reporting both quantities therefore exposes
both overall classification performance and variability across individual
experimental instances.

\begin{figure}[t]
\centering
\begin{subfigure}[t]{0.45\textwidth}
\centering
\includegraphics[width=\linewidth]{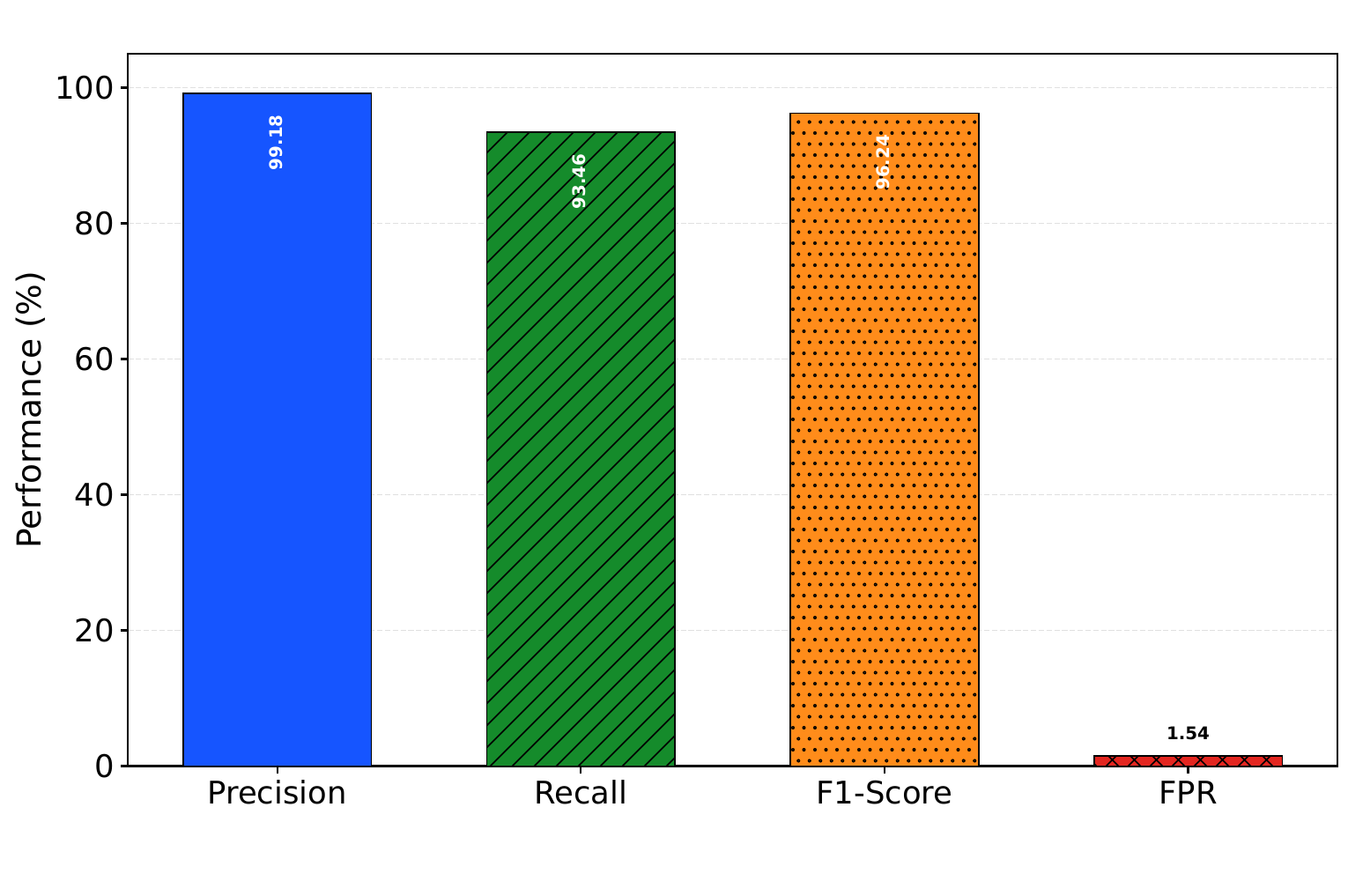}
\caption{Aggregate precision, recall, F1-score, and FPR.}
\label{fig:overall_performance_final}
\end{subfigure}\hfill
\begin{subfigure}[t]{0.45\textwidth}
\centering
\includegraphics[width=\linewidth]{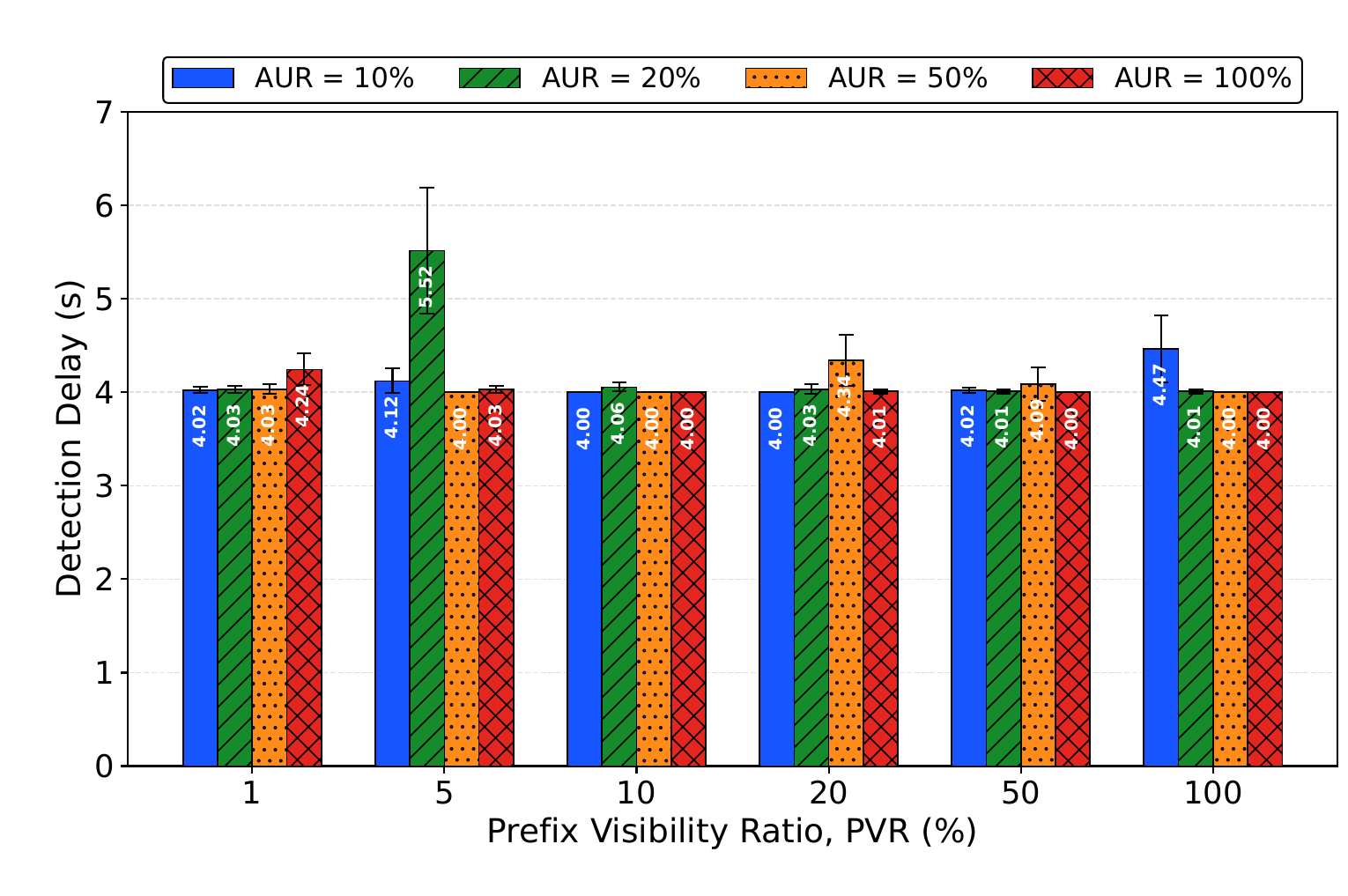}
\caption{Conditional detection delay versus PVR.}
\label{fig:delay_vs_pvr_final}
\end{subfigure}
\caption{Intrinsic VANTA performance over the independently validated 2,160-run campaign. Delay statistics include the 2,139 runs that raised at least one attack-window alert; complete-miss runs remain false negatives in the detection metrics.}
\label{fig:intrinsic_performance_final}
\end{figure}

Table~\ref{tab:aggregate_results} and
Fig.~\ref{fig:intrinsic_performance_final} summarize the aggregate results.
Detection latency is evaluated separately because a finite delay exists only
when an alert is eventually generated. Of the 2,160 completed runs, 2,139
produce a valid alert, while 21 do not. Among successfully detected runs, the
median detection delay is 4.00~s and the mean is 4.12~s, with a maximum
observed delay of 13~s. Runs without an alert remain fully accounted for as
false negatives in the classification metrics and are excluded only from the
conditional delay statistic.

\begin{table}[t]
\centering
\caption{Aggregate VANTA results over 2,160 runs.}
\label{tab:aggregate_results}
\small
\begin{tabular}{lr}
\toprule
Metric & Value \\
\midrule
True positives / false negatives & 20,187 / 1,413 \\
True negatives / false positives & 10,634 / 166 \\
Precision / Recall & 99.18\% / 93.46\% \\
Global F1-score / FPR & 96.24\% / 1.54\% \\
Median / mean detection delay & 4.00 / 4.12~s \\
Mean / median absolute PVR error & 0.0246 / 0.00077 \\
\bottomrule
\end{tabular}
\end{table}

\subsection{Matched-Observability Reference Comparison}

\begin{figure}[t]
\centering
\begin{subfigure}[t]{0.44\textwidth}
\centering
\includegraphics[width=\linewidth]{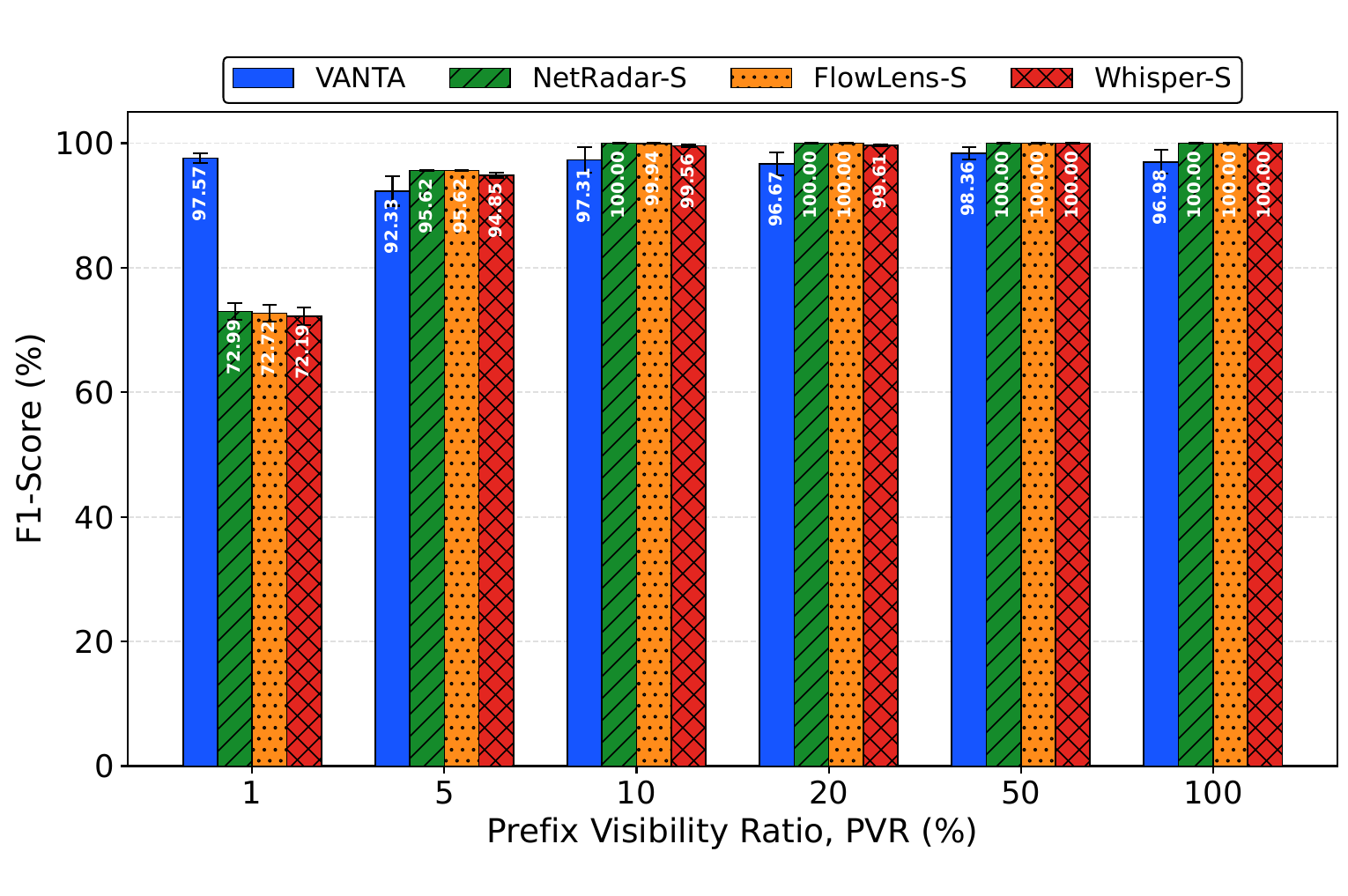}
\caption{F1-score versus PVR.}
\label{fig:comparative_f1_final}
\end{subfigure}\hfill
\begin{subfigure}[t]{0.44\textwidth}
\centering
\includegraphics[width=\linewidth]{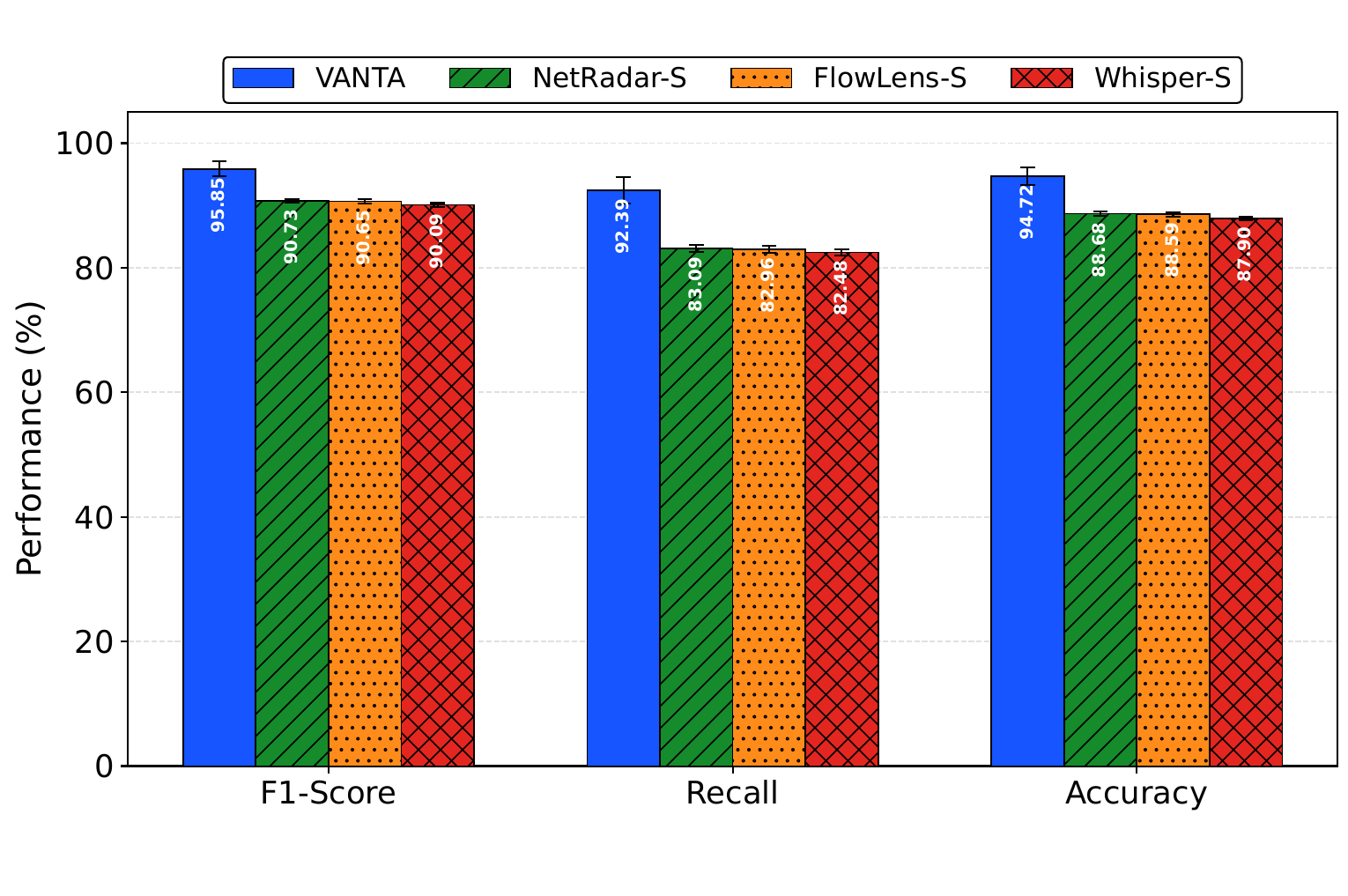}
\caption{Performance for $PVR\leq10\%$.}
\label{fig:low_visibility_final}
\end{subfigure}
\caption{Matched-observability evaluation on held-out repetitions 16--30 using task-adapted feature-level reference implementations. Bars are means of repetition-level metrics and error bars are Student-$t$ 95\% confidence intervals. VANTA uses the independently validated primary campaign; the -S references are restricted to monitored destinations.}
\label{fig:comparison_final}
\end{figure}

Figure~\ref{fig:comparison_final} compares VANTA with the three task-adapted
reference implementations when all methods are constrained by matched
destination observability. At the configured $PVR=1\%$ condition, VANTA
achieves a mean held-out F1-score of 97.57\%, compared with 72.99\% for
NetRadar-S, 72.72\% for FlowLens-S, and 72.19\% for Whisper-S. The difference
is particularly pronounced in recall: VANTA reaches 95.56\%, whereas the
strongest reference reaches 57.61\%. This result indicates that, under the
smallest configured visibility condition, the complete VANTA pipeline
preserves substantially more attack detections than the reference classifiers.
The comparison does not isolate the contribution of adaptive fusion alone.

The relative ranking changes as visibility increases. At configured
$PVR=5\%$, NetRadar-S and FlowLens-S each reach 95.62\% F1, compared with
92.33\% for VANTA. From $PVR=10\%$ upward, the reference methods rapidly
approach near-perfect classification on the synthetic packet signatures.
Across the complete held-out grid, pooled F1 is 96.66\% for VANTA, 95.57\%
for NetRadar-S, 95.53\% for FlowLens-S, and 95.20\% for Whisper-S. These
results show that the complete VANTA pipeline's comparative advantage is concentrated
primarily in the low-observability region rather than uniformly across the
entire PVR range.

When the configured low-visibility conditions $PVR\leq10\%$ are aggregated, VANTA achieves
95.89\% pooled F1 and 92.39\% recall, compared with 90.73\% F1 and 83.09\%
recall for NetRadar-S, the strongest matched reference in this regime
(Table~\ref{tab:low_visibility_results}). The corresponding improvements are
5.16 percentage points in F1 and 9.30 points in recall. VANTA's FPR is 0.63\%,
compared with 0.15\% for NetRadar-S and FlowLens-S. Thus, the improved recall
is obtained at the cost of a modest increase in false positives, illustrating
a different recall--false-alarm trade-off rather than uniformly better
classification on every metric.

The full-visibility reference results provide an important contrast: when all
active destinations are observable, the reference implementations reach
approximately 99.7\% F1. The large gap between these upper bounds and their
matched-observability performance at sparse PVR demonstrates that destination
visibility assumptions can materially affect both absolute performance and
the relative ranking of detection approaches.

These comparisons should nevertheless be interpreted within the task defined
in Section~\ref{sec:methodology}. NetRadar-S, FlowLens-S, and Whisper-S are
task-adapted reference implementations evaluated on binary prefix-level attack
presence. In particular, NetRadar's original end-to-end objective includes
victim localization. A localization failure does not imply that aggregate
gateway evidence could not reveal the presence of an attack. Consequently,
the results above establish the effect of matched destination observability
on the common binary detection task considered here. They do not establish
superiority over the original NetRadar system on its victim-localization
objective.

Finally, the smallest configured PVR values should be read together with the
finite-prefix discussion in Section~\ref{sec:effective_pvr}. Because
integer cardinalities can cause multiple nominal sparse-PVR settings to map
to the same realized overlap for small attacked sets, the configured labels
describe the experimental grid rather than guaranteeing distinct effective
PVR values in every topology.

\begin{table}[t]
\centering
\caption{Pooled results for matched observability and $PVR\leq10\%$.}
\label{tab:low_visibility_results}
\small
\begin{tabular}{lrrrrr}
\toprule
Method & Precision & Recall & F1 & FPR & Accuracy \\
\midrule
VANTA & 99.66 & \textbf{92.39} & \textbf{95.89} & 0.63 & \textbf{94.72} \\
NetRadar-S & 99.91 & 83.09 & 90.73 & 0.15 & 88.68 \\
FlowLens-S & 99.91 & 82.96 & 90.65 & 0.15 & 88.59 \\
Whisper-S & 99.24 & 82.48 & 90.09 & 1.26 & 87.90 \\
\bottomrule
\end{tabular}%
\end{table}

\subsection{Interaction between AUR and PVR}

\begin{figure}[t]
\centering
\begin{subfigure}[t]{0.44\textwidth}
\centering
\includegraphics[width=\linewidth]{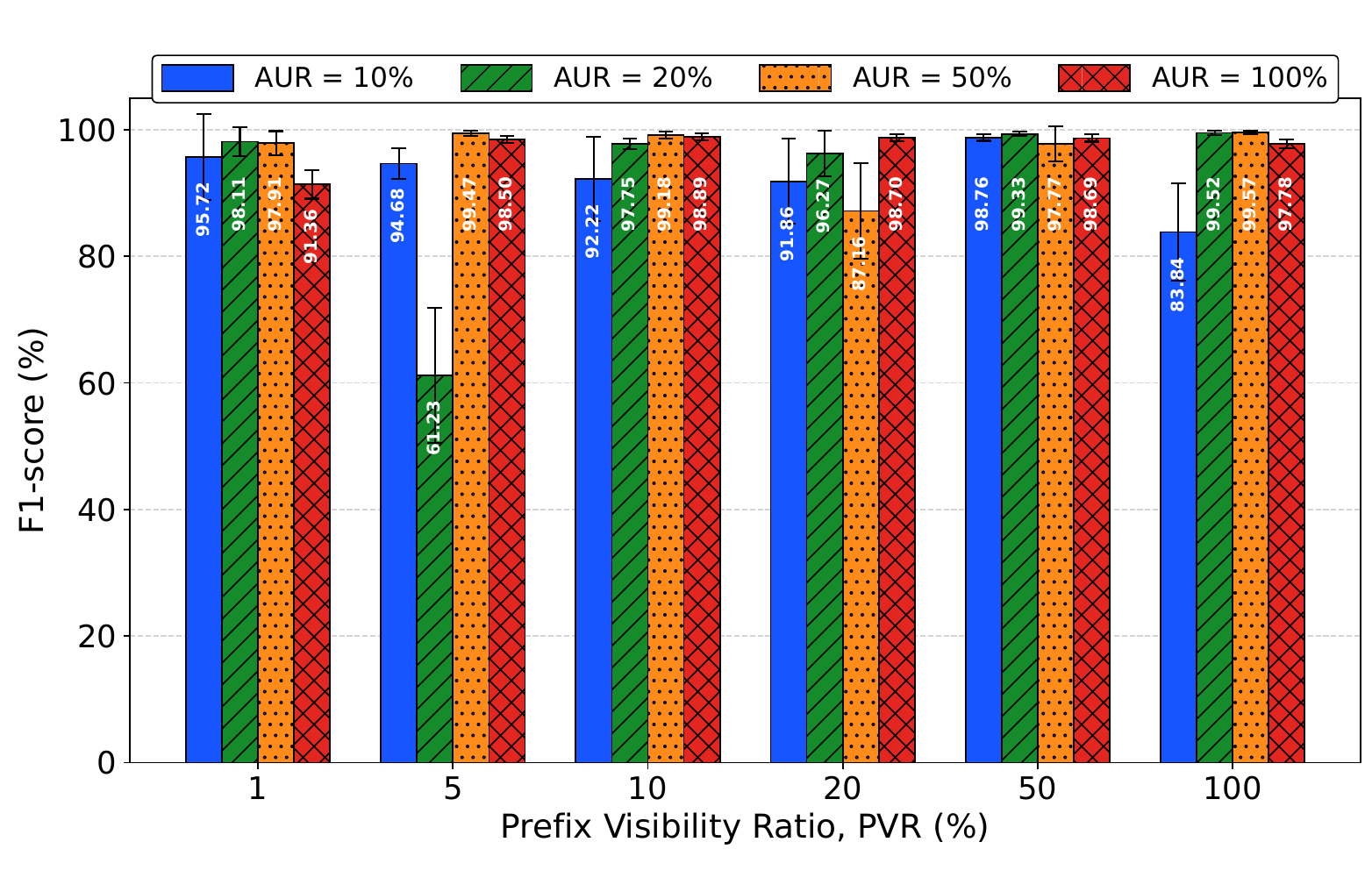}
\caption{F1-score versus PVR, separated by AUR.}
\label{fig:f1_vs_pvr_internal_final}
\end{subfigure}\hfill
\begin{subfigure}[t]{0.44\textwidth}
\centering
\includegraphics[width=\linewidth]{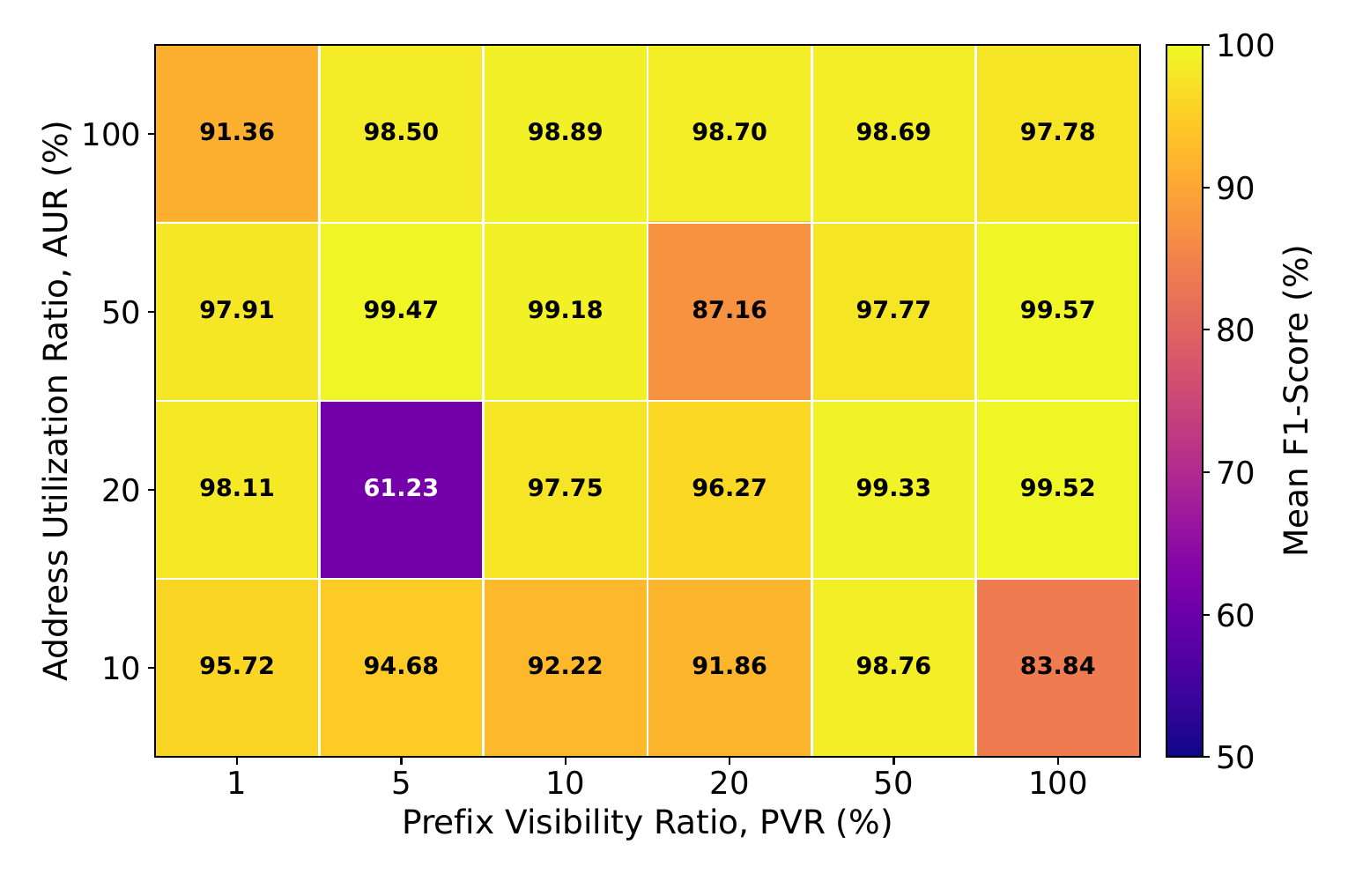}
\caption{Joint AUR--PVR effect on mean F1-score.}
\label{fig:aur_pvr_heatmap_final}
\end{subfigure}
\caption{VANTA detection performance across the complete AUR--PVR grid. Each point or cell aggregates three attack extents and 30 repetitions.}
\label{fig:observability_diagnostics_final}
\end{figure}

Figure~\ref{fig:observability_diagnostics_final} characterizes VANTA across
the complete AUR--PVR grid. Figure~\ref{fig:f1_vs_pvr_internal_final} reports
F1-score as a function of configured PVR for each AUR level, while
Fig.~\ref{fig:aur_pvr_heatmap_final} exposes the joint structure of the two
observability dimensions. Performance exceeds 0.90 in 21 of the 24
AUR--PVR cells and remains above 0.97 across most of the grid, indicating that
VANTA is generally stable over a broad range of prefix utilization and
destination visibility conditions.

The two variables affect the detector in different ways. AUR controls the
number of active destinations available in the prefix, whereas PVR determines
how much of the attacked destination set is represented in destination-specific
telemetry. Their interaction is therefore especially important at low
cardinalities, where small changes in the number of monitored or attacked
destinations can produce relatively large changes in effective visibility.

The most pronounced transition occurs around $AUR=0.2$ and configured
$PVR=0.05$. At this utilization level, 51 destinations are active and the
finite-cardinality conversion yields only a small monitored subset. Depending
on attack extent, the realized attack/monitor intersection may contain only a
single attacked destination. Host-level evidence is consequently weak or
unstable, and the decision relies more heavily on prefix-level information.

The resulting performance is not strictly monotonic in configured PVR.
Finite /24 cardinalities can contribute: neighboring nominal PVR values
can map to identical or near-identical integer overlaps, while a change of one
destination can materially alter the effective ratio. Attacked-set estimation
and the transition between host-supported and prefix-supported decisions may
also contribute. These are possible explanations, not isolated causal results.
In particular, quantization alone does not explain performance losses at full
visibility. Component ablations are needed to attribute the nonmonotonic
behavior to specific mechanisms.

\subsection{PVR Estimation Accuracy}

The online visibility estimator achieves a mean absolute error of 0.0246 and a
median absolute error of only 0.00077 with respect to the realized effective
PVR. Overall, 90.93\% of runs exhibit an estimation error below 0.05, and
96.11\% remain below 0.10. The large difference between the mean and median
indicates that estimation is highly accurate in the majority of runs, with a
small number of substantially larger errors forming a heavy tail.

\begin{figure}[t]
\centering
\includegraphics[width=0.5\columnwidth]{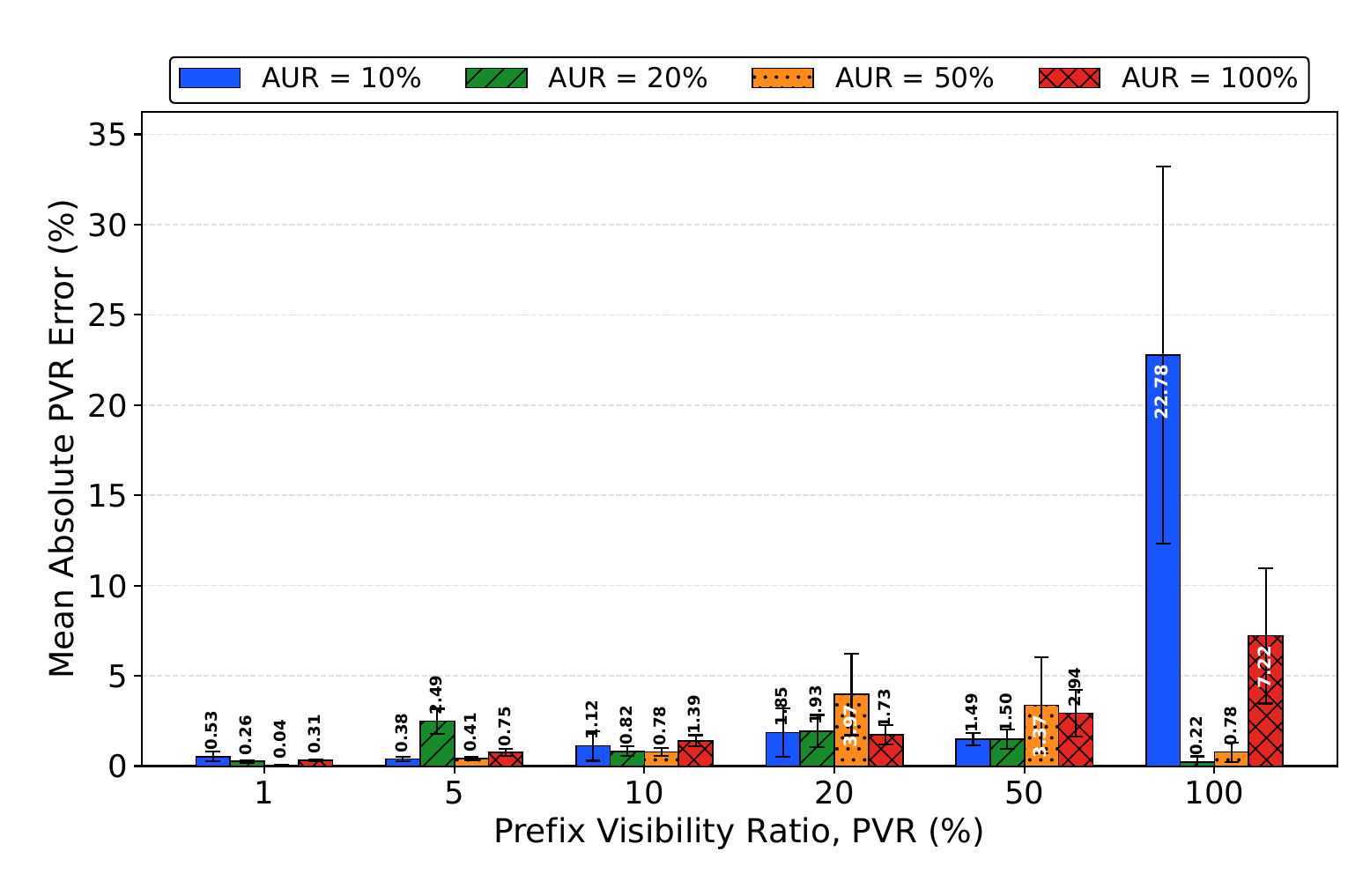}
\caption{Mean absolute PVR-estimation error versus configured PVR. Bars correspond to AUR values and error bars are Student-$t$ 95\% confidence intervals over repetition-level averages.}
\label{fig:pvr_error_final}
\end{figure}

Figure~\ref{fig:pvr_error_final} reports estimation error across the complete
AUR--PVR grid. Accuracy remains high over most observability conditions, while
the largest deviations occur when the attacked-set inference becomes incomplete
or temporarily returns an empty set. Such cases are more likely when the number
of attacked or observable destinations is small, since a single missed
destination can then produce a relatively large change in the realized ratio.

This behavior is particularly relevant because $\widehat{PVR}_t$ is not merely
a diagnostic quantity: it directly influences the weighting of host- and
prefix-level evidence. Estimation errors can therefore shift the fusion stage
toward an inappropriate visibility regime. The confidence term $c_t$ partially
mitigates this effect by reducing reliance on the PVR estimate when its
supporting evidence is weak. The error distribution consequently provides an
important diagnostic of both estimator quality and the reliability of the
subsequent adaptive-fusion decision.

\subsection{Failure Audit}

We explicitly audit the 21 runs in which VANTA fails to raise any alert during
the attack period. Among them, three topology instances, corresponding to nine
runs across the evaluated attack extents, exhibit an unusual zero-signal
pattern: the prefix score, fused score, maximum fused score, and estimated PVR
remain equal to zero throughout the attack windows. Because these runs were
completed by the experimental pipeline, we conservatively retain them in all
primary results rather than removing them post hoc.

To assess their influence, we perform a sensitivity check in which only these
nine zero-signal runs are excluded. The global F1-score increases only from
96.24\% to 96.44\%, while recall rises from 93.46\% to 93.85\%. The small
change confirms that the aggregate conclusions are not driven by these
atypical cases.

The remaining complete misses correspond to genuine detection failures rather
than uniformly absent detector signals. They are concentrated primarily in the
$AUR=0.2$, configured $PVR=0.05$ transition regime, where the realized
attack/monitor overlap can be extremely small and host-level evidence becomes
weak. This concentration is consistent with the observability analysis above:
the most difficult cases occur near the transition between host-supported and
prefix-dominant operation.

\section{Discussion}
\label{sec:discussion}

\subsection{What PVR Adds to Carpet Bombing Evaluation}

The main contribution of PVR is methodological: attacked-destination
observability constitutes a distinct experimental variable that should be
measured, controlled, and reported when evaluating carpet bombing detectors.
Traditional evaluations typically describe traffic intensity, source
population, target-set size, or monitoring deployment, but these quantities do
not directly indicate how much of the actual attacked surface is visible to the
defender.

Full-visibility evaluation can therefore conceal a detector's dependence on
destination-specific telemetry. A system may perform extremely well when every
attacked destination contributes detailed evidence, yet degrade sharply when
only a small fraction of the attacked set is observable. PVR makes this
dependence explicit by conditioning visibility on the intersection between the
attacker-selected destination set and the defender's observable set, rather
than on nominal monitoring coverage alone.

This distinction is useful beyond VANTA itself. PVR does not require a detector
to estimate visibility online or to adapt its internal decision rule. Any
carpet bombing detector can, in principle, be evaluated as a function of PVR in
order to expose its sensitivity to partial attacked-destination observability.
PVR should therefore be viewed both as an architectural input in VANTA and,
more generally, as an evaluation dimension for distributed attack detection.

\subsection{Interpreting the Multi-Level Design}

VANTA separates destination-specific and prefix-level evidence so that the
detector can remain operational when host telemetry becomes sparse. Host-level
evidence provides detailed information about monitored destinations, whereas
prefix-level evidence remains available even when the attacked/observable
overlap is small. The adaptive fusion mechanism is intended to shift reliance
between these two evidence sources as estimated visibility changes.

The present campaign demonstrates the behavior of the complete
visibility-aware pipeline and quantifies the accuracy of online PVR estimation.
It does not, however, contain the full component ablation required to isolate
the individual contribution of $S_H$, $S_P$, fixed fusion, and adaptive fusion.
In particular, strong prefix-level evidence may account for a substantial
fraction of the observed performance in low-visibility regimes.

The appropriate interpretation is therefore that the complete VANTA pipeline
remains effective over the tested observability grid, rather than that adaptive
fusion alone has been established as the unique cause of the observed gains.
A dedicated ablation comparing host-only, prefix-only, fixed-fusion, and
visibility-aware fusion variants is required to quantify the contribution of
each component independently.

\subsection{Interpreting the Reference Comparisons}

The comparative experiments address a deliberately narrower question than the
original end-to-end systems: how do representative traffic-based detection
methods behave when destination-specific inputs are restricted to the same
observable subset? NetRadar-S, FlowLens-S, and Whisper-S therefore provide
controlled reference points for the common binary prefix-level task rather than
exact reproductions of all original system objectives and architectures.

This distinction is especially important for NetRadar, whose original
end-to-end objective includes attacked-server localization. Failure to localize
a particular victim is not equivalent to failure to determine that a prefix is
under attack. Accordingly, the matched-observability results should be
interpreted as evidence that restricting destination-specific observations can
materially change binary detection behavior, not as evidence that VANTA
outperforms the original NetRadar system on victim localization.

A stronger end-to-end comparison would preserve each system's original task
while varying PVR explicitly. Such an evaluation should report both binary
attack-presence detection and victim-localization performance as visibility
decreases. It should also provide all methods with comparable gateway-level
information so that differences in performance can be attributed more cleanly
to detection design rather than to unequal access to aggregate network
evidence.

\subsection{Practical Meaning of Sparse PVR}

The largest relative differences among the evaluated methods occur in the
lowest configured visibility conditions. These results should not be
interpreted as evidence that $PVR\leq10\%$ is a typical or dominant operating
condition in production networks. PVR is attack conditional, and its
distribution depends jointly on attacker target selection and monitoring
deployment.

Sparse PVR is better interpreted here as a controlled stress condition. It can
arise when an attack targets a destination set that overlaps only weakly with
the monitored population, including cases involving inactive, unassigned,
transient, or otherwise uninstrumented addresses. In such circumstances,
nominal monitoring coverage may substantially overstate the amount of
destination-specific attack evidence actually available to the detector.

The operational distribution of PVR is currently unknown and cannot be inferred
from monitoring coverage alone. Measuring PVR in production environments,
across different address spaces and attack classes, is therefore an important
research problem in its own right. Such measurements would help determine which
visibility regimes are most representative of operational deployments.

\subsection{Estimator Error and Decision Sensitivity}

The online estimator achieves a mean absolute error of 0.0246 and a median
absolute error of 0.00077 in the validated campaign, indicating accurate
visibility estimation in most runs with a small number of larger deviations.
Because $\widehat{PVR}_t$ directly influences the fusion weight, however,
estimation error can propagate into the final detector by shifting the balance
between host- and prefix-level evidence.

The current results quantify estimator accuracy but do not isolate its precise
effect on classification performance. An oracle-PVR experiment, in which the
fusion stage receives the true effective PVR, would provide an upper bound on
the performance loss attributable specifically to estimation error. Conversely,
controlled perturbation of $\widehat{PVR}_t$ would reveal how sensitive the
decision process is to over- or under-estimation.

A related question concerns parameter sensitivity. The validated values of
$\tau_L$, $\tau_H$, estimator parameters, and decision thresholds are reported
explicitly in Table~\ref{tab:key_parameters}, but the present campaign does not
establish that these values are optimal or universally robust. A systematic
sensitivity analysis is therefore required before drawing conclusions about
the stability of the fusion mechanism across different deployment conditions.

\subsection{Limitations and Next Experimental Stage}

The current evaluation intentionally favors experimental control over
deployment realism. It uses a Mininet /24 environment with synthetic traffic,
50 legitimate clients, and 100 distributed attack sources. This setting allows
AUR and attack extent to be varied across a controlled grid of
attacked/observable overlap. Monitoring coverage is coupled to the configured
PVR, and the evaluated targets are active destinations. The setting does not
reproduce ISP-scale routing diversity,
application behavior, background-traffic variability, or operational
measurement noise.

The comparative reference implementations introduce additional scope
limitations. The ``-S'' methods are transparent feature-level adaptations
rather than exact executions of the original end-to-end systems. Moreover,
although VANTA and the comparative campaign use the same configuration and seed
schedule, they originate from independent experimental runs and are therefore
seed-aligned rather than packet-for-packet paired. The synthetic generator also
creates comparatively regular attack signatures, as reflected by the
near-perfect performance of the reference classifiers under full visibility.

The next experimental stage should consequently extend both realism and causal
analysis. In particular, it should include:
(i) $S_H$-only, $S_P$-only, fixed-fusion, and adaptive-fusion ablations,
(ii) oracle versus estimated PVR,
(iii) sensitivity analysis for $\tau_L$, $\tau_H$, and the decision thresholds,
(iv) paired executions in which all methods receive the same traffic traces and
equivalent gateway evidence,
(v) original-task evaluation of victim localization as PVR decreases,
(vi) larger target prefixes to reduce sparse-count quantization,
(vii) adaptive attackers capable of rotating destination sets according to
inferred monitoring coverage, and
(viii) replay or traffic generation grounded in real operational traces.

These extensions are necessary before making stronger claims about end-to-end
system superiority, large-scale deployment behavior, or the prevalence of
specific PVR regimes. The present results should instead be interpreted as a
controlled demonstration that attacked-destination observability is a
measurable and consequential dimension of carpet bombing detection.

\section{Conclusion}
\label{sec:conclusion}

This paper introduced the Prefix Visibility Ratio (PVR) as an explicit metric
for attacked-destination observability in carpet bombing detection. Unlike
address utilization or nominal monitoring coverage, PVR is conditioned on the
destinations actually selected by the attacker and therefore captures the
fraction of the attacked surface for which destination-specific evidence is
available to the defender.

We used VANTA as a concrete visibility-aware detection framework that estimates
PVR online, maintains separate host-level and prefix-level anomaly scores, and
adapts their fusion according to both estimated visibility and estimator
confidence. Across 2,160 validated Mininet runs, VANTA achieves 99.18\%
precision, 93.46\% recall, and a pooled F1-score of 96.24\%, while maintaining
a mean detection delay of 4.12~s among successfully detected runs.

A separate matched-observability study with task-adapted reference
implementations further shows that binary detection behavior can change
substantially as destination-specific visibility decreases. The largest
differences appear in the sparsest configured regimes, whereas the reference
methods approach near-perfect classification under full visibility. These
results should be interpreted as evidence that observability assumptions can
materially affect detector performance and relative ranking, rather than as an
end-to-end reproduction of the original referenced systems or their victim-localization objectives.

The broader contribution is therefore an evaluation principle as much as a
detector design. Carpet bombing defenses should report not only traffic
intensity, attack extent, and address utilization, but also the
attacked/observable overlap that determines how much destination-specific
evidence is actually available. Making this quantity explicit exposes hidden
visibility assumptions and provides a basis for more rigorous evaluation of
detection and localization under partial telemetry.

Future work should extend this analysis to larger prefixes, real operational
traffic, adaptive monitor-aware attackers, component ablations, oracle-PVR
comparisons, and original-task evaluation of victim localization. Such studies
will be necessary to determine how the PVR distributions observed in controlled
experiments translate to production environments and to quantify more precisely
which components of visibility-aware detection provide the greatest benefit.

\section*{Artifact Availability}




The implementation, experimental configurations, task-adapted
reference implementations, analysis and verification scripts,
and archived result files are publicly available on Zenodo
as version \texttt{v1.0-arxiv}:
\textcolor{blue}{\url{https://doi.org/10.5281/zenodo.23534795}}.

This version provides the material needed to verify the principal
numerical results and regenerate the corresponding figures.
Subsequent experimental extensions are excluded from this archive
and from the results reported in this paper.

\input{VANTA.bbl}

\end{document}

%% file: VANTA.bbl

%% file: main.bbl
\begin{thebibliography}{10}
\providecommand{\url}[1]{#1}
\csname url@samestyle\endcsname
\providecommand{\newblock}{\relax}
\providecommand{\bibinfo}[2]{#2}
\providecommand{\BIBentrySTDinterwordspacing}{\spaceskip=0pt\relax}
\providecommand{\BIBentryALTinterwordstretchfactor}{4}
\providecommand{\BIBentryALTinterwordspacing}{\spaceskip=\fontdimen2\font plus
\BIBentryALTinterwordstretchfactor\fontdimen3\font minus
  \fontdimen4\font\relax}
\providecommand{\BIBforeignlanguage}[2]{{%
\expandafter\ifx\csname l@#1\endcsname\relax
\typeout{** WARNING: IEEEtran.bst: No hyphenation pattern has been}%
\typeout{** loaded for the language `#1'. Using the pattern for}%
\typeout{** the default language instead.}%
\else
\language=\csname l@#1\endcsname
\fi
#2}}
\providecommand{\BIBdecl}{\relax}
\BIBdecl

\bibitem{mirkovic2004taxonomy}
J.~Mirkovic and P.~Reiher, ``A taxonomy of ddos attack and ddos defense
  mechanisms,'' \emph{ACM SIGCOMM Computer Communication Review}, vol.~34,
  no.~2, pp. 39--53, 2004.

\bibitem{zargar2013survey}
S.~T. Zargar, J.~Joshi, and D.~Tipper, ``A survey of defense mechanisms against
  distributed denial of service (ddos) flooding attacks,'' \emph{IEEE
  Communications Surveys \& Tutorials}, vol.~15, no.~4, pp. 2046--2069, 2013.

\bibitem{heinrich2021carpetbombing}
\BIBentryALTinterwordspacing
T.~Heinrich, R.~R. Obelheiro, and C.~A. Maziero, ``New kids on the drdos block:
  Characterizing multiprotocol and carpet bombing attacks,'' in \emph{Passive
  and Active Measurement: 22nd International Conference, PAM 2021, Virtual
  Event, March 29 – April 1, 2021, Proceedings}.\hskip 1em plus 0.5em minus
  0.4em\relax Berlin, Heidelberg: Springer-Verlag, 2021, p. 269–283.
  [Online]. Available: \url{https://doi.org/10.1007/978-3-030-72582-2_16}
\BIBentrySTDinterwordspacing

\bibitem{huici2007edge}
F.~Huici and M.~Handley, ``An edge-to-edge filtering architecture against
  dos,'' \emph{ACM SIGCOMM Computer Communication Review}, vol.~37, no.~2, pp.
  39--50, 2007.

\bibitem{wagner2021united}
D.~Wagner, D.~Kopp, M.~Wichtlhuber, C.~Dietzel, O.~Hohlfeld, G.~Smaragdakis,
  and A.~Feldmann, ``United we stand: Collaborative detection and mitigation of
  amplification ddos attacks at scale,'' in \emph{Proceedings of the 2021 ACM
  SIGSAC conference on computer and communications security}, 2021, pp.
  970--987.

\bibitem{netradar}
J.~Pan, L.~Zhang, X.~Si, J.~Zhang, X.~Zhang, and Y.~Cui, ``Netradar: Enabling
  robust carpet bombing ddos detection,'' in \emph{Proceedings of the Network
  and Distributed System Security Symposium (NDSS)}, 01 2026.

\bibitem{flowlens}
D.~Barradas, N.~Santos, L.~Rodrigues, S.~Signorello, F.~M.~V. Ramos, and
  A.~Madeira, ``Flowlens: Enabling efficient flow classification for ml-based
  network security applications,'' in \emph{Proceedings of the Network and
  Distributed System Security Symposium (NDSS)}, 2021, pp. 1--16.

\bibitem{whisper}
C.~Fu, Q.~Li, M.~Shen, and K.~Xu, ``Realtime robust malicious traffic detection
  via frequency domain analysis,'' in \emph{Proceedings of the 2021 ACM SIGSAC
  Conference on Computer and Communications Security}, 2021, pp. 3431--3446.

\bibitem{richter2015surveying}
P.~Richter, G.~Smaragdakis, D.~Plonka, and A.~Berger, ``Beyond counting: New
  perspectives on the active IPv4 address space,'' in \emph{Proceedings of ACM
  Internet Measurement Conference (IMC)}, 2016.

\bibitem{caida}
{CAIDA}, ``Caida ucsd anonymized internet traces,'' 2025,
  \url{https://www.caida.org} (accessed June 2026).

\bibitem{liu2016middlepolice}
Z.~Liu, H.~Jin, Y.-C. Hu, and M.~Bailey, ``Middlepolice: Toward enforcing
  destination-defined policies in the middle of the internet,'' in
  \emph{Proceedings of the 2016 ACM SIGSAC Conference on Computer and
  Communications Security}, 2016, pp. 1268--1279.

\bibitem{hu2025sistar}
J.~Hu, F.~Guo, Q.~Zhou, Y.~Zhang, Z.~Gao, Y.~Han, and Z.~Lv, ``Sistar: An
  efficient ddos detection and mitigation framework utilizing programmable data
  planes,'' in \emph{Proceedings of the 2025 ACM SIGSAC Conference on Computer
  and Communications Security}, 2025, pp. 4589--4603.

\bibitem{bailey2005network}
M.~Bailey, E.~Cooke, F.~Jahanian, D.~Watson, and J.~Nazario, ``The internet
  motion sensor: A distributed blackhole monitoring system,'' in
  \emph{Proceedings of NDSS}, 2005.

\bibitem{moore2006inferring}
D.~Moore, C.~Shannon, D.~J. Brown, G.~M. Voelker, and S.~Savage, ``Inferring
  internet denial-of-service activity,'' \emph{ACM Transactions on Computer
  Systems}, vol.~24, no.~2, pp. 115--139, 2006.

\bibitem{sommese2022investigating}
R.~Sommese, K.~Claffy, R.~van Rijswijk-Deij, A.~Chattopadhyay, A.~Dainotti,
  A.~Sperotto, and M.~Jonker, ``Investigating the impact of ddos attacks on dns
  infrastructure,'' in \emph{proceedings of the 22nd ACM Internet Measurement
  Conference}, 2022, pp. 51--64.

\bibitem{nawrocki2021far}
M.~Nawrocki, M.~Jonker, T.~C. Schmidt, and M.~W{\"a}hlisch, ``The far side of
  dns amplification: tracing the ddos attack ecosystem from the internet
  core,'' in \emph{Proceedings of the 21st ACM Internet Measurement
  Conference}, 2021, pp. 419--434.

\bibitem{lakhina2004diagnosing}
A.~Lakhina, M.~Crovella, and C.~Diot, ``Diagnosing network-wide traffic
  anomalies,'' in \emph{Proceedings of ACM SIGCOMM}, 2004, pp. 219--230.

\bibitem{barford2002signal}
P.~Barford, J.~Kline, D.~Plonka, and A.~Ron, ``A signal analysis of network
  traffic anomalies,'' in \emph{Proceedings of ACM SIGCOMM Internet Measurement
  Workshop}, 2002.

\bibitem{krishnamurthy2003sketch}
B.~Krishnamurthy, S.~Sen, Y.~Zhang, and Y.~Chen, ``Sketch-based change
  detection: Methods, evaluation, and applications,'' in \emph{Proceedings of
  ACM Internet Measurement Conference (IMC)}, 2003, pp. 234--247.

\bibitem{alcoz2022aggregate}
A.~G. Alcoz, M.~Strohmeier, V.~Lenders, and L.~Vanbever, ``Aggregate-based
  congestion control for pulse-wave ddos defense,'' in \emph{Proceedings of the
  ACM SIGCOMM 2022 Conference}, 2022, pp. 693--706.

\bibitem{garcia2009anomaly}
P.~Garcia-Teodoro, J.~Diaz-Verdejo, G.~Macia-Fernandez, and E.~Vazquez,
  ``Anomaly-based network intrusion detection: Techniques, systems and
  challenges,'' \emph{Computers \& Security}, vol.~28, no. 1-2, pp. 18--28,
  2009.

\bibitem{bhuyan2014network}
M.~H. Bhuyan, D.~K. Bhattacharyya, and J.~K. Kalita, ``Network anomaly
  detection: Methods, systems and tools,'' \emph{IEEE Communications Surveys \&
  Tutorials}, vol.~16, no.~1, pp. 303--336, 2014.

\bibitem{sommer2010outside}
R.~Sommer and V.~Paxson, ``Outside the closed world: On using machine learning
  for network intrusion detection,'' in \emph{IEEE Symposium on Security and
  Privacy}, 2010, pp. 305--316.

\bibitem{estan2002newdirections}
C.~Estan and G.~Varghese, ``New directions in traffic measurement and
  accounting,'' in \emph{Proceedings of ACM SIGCOMM}, 2002, pp. 323--336.

\bibitem{claise2008ipfix}
B.~Claise, ``Specification of the ip flow information export (ipfix)
  protocol,'' IETF, Tech. Rep. RFC 5101, 2008.

\bibitem{kompella2014every}
R.~R. Kompella, K.~Levchenko, A.~C. Snoeren, and G.~Varghese, ``Every
  microsecond counts: Tracking fine-grain latencies with a lossy difference
  aggregator,'' in \emph{Proceedings of ACM SIGCOMM}, 2009, pp. 255--266.

\bibitem{feamster2014road}
N.~Feamster, J.~Rexford, and E.~Zegura, ``The road to SDN: An intellectual
  history of programmable networks,'' \emph{Queue}, vol.~11, no.~12, 2013.

\bibitem{montgomery2019spc}
D.~C. Montgomery, \emph{Introduction to Statistical Quality Control},
  8th~ed.\hskip 1em plus 0.5em minus 0.4em\relax John Wiley \& Sons, 2019.

\end{thebibliography}
